%% file: ms.tex
\documentclass[twocolumn]{aastex63}

\usepackage{gensymb}

\newcommand{\mbh}{{\mbox{$M_\mathrm{BH}$}}}
\newcommand{\mmol}{{\mbox{$M_\mathrm{H_2}$}}}

\newcommand{\coline}{{\mbox{CO(2$-$1)}}}
\newcommand{\lco}{{\mbox{$L^\prime_\mathrm{CO}$}}}
\newcommand{\lir}{{\mbox{$L_\mathrm{IR}$}}}
\newcommand{\lagn}{{\mbox{$\lambda L_\lambda(5100\,\mathrm{\AA})$}}}
\newcommand{\aco}{{\mbox{$\alpha_\mathrm{CO}$}}}
\newcommand{\uaco}{{\mbox{$M_\odot\,\mathrm{(K\,km\,s^{-1}\,pc^{2})^{-1}}$}}}
\newcommand{\comol}{{CO-to-H$_2$}}
\newcommand{\kms}{{\mbox{$\mathrm{km\,s^{-1}}$}}}
\newcommand{\ergs}{{\mbox{$\mathrm{erg\,s^{-1}}$}}}
\newcommand{\usfr}{{\mbox{$M_\odot\,\mathrm{yr}^{-1}$}}}

\font\sevenrm=cmr7 scaled 1000
\newcommand{\ci}{[C{\sevenrm\,I}]}
\newcommand{\hi}{H{\sevenrm\,I}}

\newcommand{\hst}{{HST}}

\newcommand{\tnma}{{\tablenotemark{\tiny a}}}
\newcommand{\mcl}{\multicolumn}
\newcommand{\ph}{\phantom}

\newcommand{\linmix}{\texttt{Linmix}}
\newcommand{\galfit}{\texttt{GALFIT}}

\graphicspath{{./}{figures/}}

\submitjournal{ApJ}

\shorttitle{Molecular Gas in Quasar Host Galaxies}
\shortauthors{Shangguan et al.}

\begin{document}

\title{AGN Feedback and Star Formation of Quasar Host Galaxies: Insights from the Molecular Gas}

\correspondingauthor{Jinyi Shangguan}
\email{shangguan@mpe.mpg.de}

\author[0000-0002-4569-9009]{Jinyi Shangguan}
\affil{Max-Planck-Institut f\"{u}r Extraterrestrische Physik (MPE), 
Giessenbachstr., D-85748 Garching, Germany}
\affiliation{Kavli Institute for Astronomy and Astrophysics, Peking University,
Beijing 100871, China}

\author[0000-0001-6947-5846]{Luis C. Ho}
\affil{Kavli Institute for Astronomy and Astrophysics, Peking University,
Beijing 100871, China}
\affiliation{Department of Astronomy, School of Physics, Peking University,
Beijing 100871, China}

\author[0000-0002-8686-8737]{Franz E. Bauer}
\affil{Instituto de Astrof{\'{\i}}sica and Centro de Astroingenier{\'{\i}}a, Facultad 
de F{\'{i}}sica, Pontificia Universidad Cat{\'{o}}lica de Chile, Casilla 306, 
Santiago 22, Chile} 
\affiliation{Millennium Institute of Astrophysics (MAS), Nuncio Monse{\~{n}}or 
S{\'{o}}tero Sanz 100, Providencia, Santiago, Chile} 
\affiliation{Space Science Institute, 4750 Walnut Street, Suite 205, Boulder, 
Colorado 80301}

\author[0000-0003-4956-5742]{Ran Wang}
\affil{Kavli Institute for Astronomy and Astrophysics, Peking University,
Beijing 100871, China}
\affiliation{Department of Astronomy, School of Physics, Peking University,
Beijing 100871, China}

\author[0000-0001-7568-6412]{Ezequiel Treister}
\affil{Instituto de Astrof{\'{\i}}sica and Centro de Astroingenier{\'{\i}}a, Facultad 
de F{\'{i}}sica, Pontificia Universidad Cat{\'{o}}lica de Chile, Casilla 306, 
Santiago 22, Chile} 

\begin{abstract}

The molecular gas serves as a key probe of the complex interplay between black hole accretion and star formation in the host galaxies of active galactic nuclei (AGNs).  We use CO(2--1) observations from a new ALMA survey, in conjunction with literature measurements, to investigate the molecular gas properties of a representative sample of 40 $z < 0.3$  Palomar-Green quasars, the largest and most sensitive study of molecular gas emission to date for nearby quasars.  We find that the AGN luminosity correlates with both the CO luminosity and black hole mass, suggesting that AGN activity is loosely coupled to the cold gas reservoir of the host.  The observed strong correlation between host galaxy total infrared luminosity and AGN luminosity arises from their common dependence on the molecular gas.  We argue that the total infrared luminosity, at least for low-redshift quasars, can be used to derive reliable star formation rates for the host galaxy.  The host galaxies of low-redshift quasars have molecular gas content similar to that of star-forming galaxies of comparable stellar mass.  Moreover, they share similar gas kinematics, as evidenced by their CO Tully-Fisher relation and the absence of detectable molecular outflows down to sensitive limits.  There is no sign that AGN feedback quenches star formation for the quasars in our sample.  On the contrary, the abundant gas supply forms stars prodigiously, at a rate that places most of them above the star-forming main sequence and with an efficiency that rivals that of starburst systems. 
\end{abstract}


\section{Introduction} 
\label{sec:intro}

It is still debated whether, when, and how supermassive black holes (BHs) coevolve with galaxies \citep{Kormendy2013ARAA,Heckman2014ARAA,Greene2020}.  Feedback from active galactic nuclei (AGNs) appears to be a key ingredient for linking the central BH to its host galaxy, by shutting off its star formation and maintaining its quiescence after it has been quenched \citep{Fabian2012ARAA}.   However, the detailed physical mechanism by which AGN feedback operates remains unclear.  Radiation pressure from quasar-mode feedback can blow the gas out efficiently from the host galaxy \citep{Silk1998AA,Harrison2018NatAs}, but the contribution of mechanical feedback by AGN-driven winds cannot be neglected \citep{Yuan2018}.  Both numerical simulations (e.g., \citealt{DiMatteo2005Natur, Hopkins2016MNRAS,Costa2018MNRAS, Barnes2020}) and observations (e.g., \citealt{Cicone2014AA, Zakamska2014MNRAS, Perna2015AA,Morganti2016AA,Fiore2017AA,Nesvadba2017AA, Baron2018MNRAS,Fluetsch2019MNRAS,ForsterSchreiber2019, HerreraCamus2019ApJ}) show clear evidence that AGNs can launch strong multi-phase gas outflows \citep{Cicone2018NatAs}.  Meanwhile, it is still unclear the degree to which an AGN truly expels cold gas from its host galaxy and curtails star formation.  The detection of large amounts of cold gas (atomic and molecular) in AGN host galaxies casts doubt on the overall effectiveness of quasar-mode feedback \citep{Ho2008ApJ, Konig2009, Fabello2011, Gereb2015, Zhu2015, Shangguan2018ApJ,Shangguan2020ApJS, Ellison2019MNRAS, Shangguan2019ApJ, Yesuf2020}, at least in relatively massive systems \citep{Bradford2018}.

Spatially resolved spectroscopic observations reveal that AGN-driven outflows can actually trigger star formation, presumably by compressing the cold interstellar medium in the host galaxy \citep{Cresci2015ApJ,Carniani2016AA,Maiolino2017Natur, Gallagher2019MNRAS}.  Instead of suppressing star formation through the aforementioned ``negative'' feedback, an AGN may impart ``positive'' feedback and enhance star formation in a galaxy \citep{Cresci2018NatAs}.  The relative balance of these two opposing effects remains elusive, especially for high-redshift ($z \gtrsim 1$) AGNs (e.g., \citealt{Scholtz2020MNRAS}).  Indeed, BHs and their host galaxies perhaps coevolve without much mediation from AGN feedback at all \citep{Kormendy2013ARAA}.  Dissipative mergers at early times (e.g., $z \gtrsim 2$) drive efficient gas flows that rapidly grow central BHs and galactic bulges.  Once an initial correlation is established between BH mass and bulge mass, no matter how loosely, mass averaging by galaxy mergers inevitably generates a tight linear correlation involving the two quantities \citep{Peng2007ApJ,Hirschmann2010MNRAS,Jahnke2011ApJ}.

A closely related issue involves the connection between BH accretion and star formation in the host galaxy.  Here, too, no consensus has yet emerged.  While some contend that star formation rate (SFR) correlates strongly with AGN luminosity across a wide range of redshifts, implicating the simultaneous growth of the BH and its host (e.g., \citealt{Bonfield2011MNRAS,Rosario2012AA,Xu2015ApJ,Lanzuisi2017AA,Dai2018MNRAS,Stemo2020ApJ}), others find only moderate or no correlation \citep{Rosario2013AA,Azadi2015ApJ,Stanley2015MNRAS,Stanley2017MNRAS}.  The disagreement is partly explained by the different timescales of the measured SFR and AGN variability \citep{Hickox2014ApJ}.  At high redshifts (e.g., $z>1$), many works do not find significant correlation between SFR and AGN luminosity (e.g., \citealt{Shao2010AA,Rosario2012AA,Schulze2019MNRAS}; but see \citealt{Stemo2020ApJ}).  For low redshifts, the correlation is moderate in weak AGNs (e.g., \citealt{Shimizu2017MNRAS}) but strong in more powerful systems (e.g., \citealt{Netzer2009MNRAS,Imanishi2011PASJ,Xia2012ApJ,Zhuang2020}).  It has also been argued that the relationship between SFR and AGN luminosity is not intrinsic but instead arises indirectly from the the mutual correlation of both quantities to the host galaxy stellar mass (e.g., \citealt{Xu2015ApJ,Yang2017ApJ,Suh2019ApJ,Stemo2020ApJ}), stellar mass surface density \citep{Ni2020}, or BH mass (e.g., \citealt{Stanley2017MNRAS}).  

Using a hierarchical Bayesian model, \cite{Grimmett2020arXiv} recently demonstrated that AGNs of higher X-ray luminosity more tightly correlate with higher SFRs than lower luminosity AGNs.  In any event, when present, the association between star formation and BH accretion seems to occur preferentially on nuclear scales, both in moderately luminous AGNs (\citealt{Davies2007ApJ,Watabe2008ApJ,DiamondStanic2012ApJ, Esquej2014ApJ, Zhuang2020}) and in more powerful quasars (\citealt{Imanishi2011PASJ, Canalizo2013ApJ,Bessiere2014MNRAS,Bessiere2017MNRAS,Kim2019ApJ, Zhao2019ApJ}).  This is not unexpected, as cold gas on small scales naturally fuels nuclear star formation and feeds the BH (\citealt{Hopkins2010MNRAS,Volonteri2015MNRAS,Gan2019ApJ}).  Observations of nearby galactic nuclei offer useful clues.  For example, the molecular gas in the central regions of many galaxies (e.g., \citealt{Scoville1994,Smith1996ApJ,Rubin1997AJ,Hicks2009ApJ,GarciaBurillo2005AA,GarciaBurillo2019AA,Izumi2018ApJ,Salak2018ApJ,Treister2018ApJ,Boizelle2019ApJ}), including the Milky Way (e.g., \citealt{Ho1991, Hsieh2017}), resides in the form of a circumnuclear disk.  Theoretical works suggest that the star formation in the circumnuclear disk may control the gas fuelling BH accretion (\citealt{Shlosman1989Natur, Goodman2003MNRAS,Thompson2005ApJ,Kawakatu2008ApJ, Vollmer2008AA,Chamani2017AA,Kawakatu2020ApJ}).  Quasars, by virtue of their greater distances, lack detailed observations of their cold gas reservoirs on such small physical scales.  Moreover, if quasars preferentially experience violent dissipative processes, such as major galaxy mergers \citep{Treister2012ApJ}, their internal gas kinematics may be more complicated than in nearby AGNs.

Although submillimeter and radio spectral-line observations of quasars are time-consuming, the number of quasars with molecular line detections is rapidly expanding, both at low ($z \lesssim 1$; \citealt{Evans2001AJ,Evans2006AJ,Bertram2007AA, Krips2012, Xia2012ApJ, VillarMartin2013MNRAS, Rodriguez2014, Shangguan2020ApJS}) and high ($z \gtrsim 2$; \citealt{Carilli2002AJ,Walter2004ApJ,Riechers2006ApJ, Wang2013ApJ,Wang2016ApJ,Shao2017ApJ}) redshifts.  In their study of star formation and molecular gas in low-redshift quasars, \cite{Husemann2017MNRAS} found that the gas fraction and star formation efficiency ($\mathrm{SFE} \equiv \mathrm{SFR}/\mmol$) of quasar host galaxies are related to their galaxy morphology.  A trend between AGN luminosity and molecular gas mass further suggested that accretion power may be linked directly with the circumnuclear gas reservoir, but the heterogeneous nature of their sample and the inclusion of many non-detections preclude definitive conclusions to be drawn.  

\cite{Shangguan2020ApJS} recently completed a new CO(2--1) survey of 23 $z<0.1$ Palomar-Green (PG; \citealt{Schmidt1983ApJ}) quasars with the Atacama Large Millimeter/submillimeter Array (ALMA).  Together with CO(1--0) data for additional sources in the literature, they obtained CO measurements with a detection rate of 83\% for 40 quasars that form a representative subset of the entire sample of PG quasars with $z<0.3$.  The \lir--\lco\ relation of the quasar host galaxies follows the relation known for starburst galaxies, while their CO line ratios ($L^\prime_{\scriptsize \mbox{CO(2--1)}}/L^\prime_{\scriptsize \mbox{CO(1--0)}}$) and the CO-to-H$_2$ conversion factors resemble those of star-forming galaxies.  

As a follow-up to \cite{Shangguan2020ApJS}, the main goals of the current paper are to search for evidence of quasar-mode AGN feedback and to investigate the AGN-starburst connection in quasar host galaxies.  The paper is organized as follows:  The sample and measurements are summarized in Section~\ref{sec:smp}.  Section~\ref{sec:res} analyzes the molecular gas masses and kinematics, correlations between CO luminosities and AGN properties, and constraints on mass outflow rates.  Section~\ref{sec:disc} discusses the relationship between molecular gas masses and AGN and star formation properties, and their implications for BH--galaxy coevolution.  A summary is given in Section~\ref{sec:sum}.  This work adopts the following parameters for a $\Lambda$CDM cosmology: $\Omega_m = 0.308$, $\Omega_\Lambda = 0.692$, and 
$H_{0}=67.8$ km s$^{-1}$ Mpc$^{-1}$ \citep{Planck2016AA}.

\section{Sample and Data Analysis} 
\label{sec:smp}

\cite{Shangguan2020ApJS} reduced the ALMA Atacama Compact Array (ACA) data using the Common Astronomy Software Application\footnote{\url{https://casa.nrao.edu}} (CASA; \citealt{McMullin2007ASP}).  We successfully detected $^{12}$CO(2--1) emission in 21 out of the 23 PG quasars observed.  The CO line flux was measured from the intensity map within the $2\,\sigma$ contour of the source emission.  Spectra were extracted from the data cube based on the same aperture.  We measured the CO line width by fitting the emission line with a Gaussian double-peak function as in \cite{Tiley2016MNRAS}.  Combining with CO(1--0) measurements published in the literature, we assembled in total 40 $z<0.3$ PG~quasars having CO line observations.  This is the largest sample of low-redshift quasars with CO data to date, and it forms a representative subset of all (70) PG quasars within this redshift range.  Compared to the literature sample, the new ALMA observations extend the parameter space probed by 0.5--1.0 dex in most parameters studied here.  Using the 15 objects with both CO(1--0) and CO(2--1) measurements, among them eight detected in both lines, we derived a CO line ratio of CO(2--1)/CO(1--0) = $0.62^{+0.15}_{-0.07}$, which enabled us to convert all CO line luminosities from $L_{\scriptsize \mbox{CO(2--1)}}$ to $L_{\scriptsize \mbox{CO(1--0)}}$.  We calculate molecular gas masses assuming a \comol\ conversion factor of $\alpha_\mathrm{CO} = 3.1$ \uaco, as recommended by \cite{Sandstrom2013ApJ} for nearby star-forming galaxies, since this value shows reasonably good consistency with gas masses predicted from dust emission \citep{Shangguan2018ApJ}.

\cite{Shangguan2020ApJS} compiled the AGN 5100~\AA\ continuum luminosity [\lagn], the BH mass (\mbh), as well as the stellar mass ($M_*$) and infrared (IR) luminosity (\lir) of the host galaxy from Table~1 of \cite{Shangguan2018ApJ}.  The BH mass is estimated from the broad H$\beta$ full width at half maximum and 5100~\AA\ continuum luminosity of the AGN, using the so-called single-epoch method \citep{Ho2015ApJ}.  \cite{Shangguan2018ApJ} modelled the $\sim$ 1--500 \micron\ global IR spectral energy distribution of the $z<0.5$ PG quasars with a combination of models for stellar emission, hot dust emission of the AGN ``torus'', cold dust emission from the host galaxy, and, if needed, radio jet. The IR luminosity of the host galaxy is derived by integrating the 8--1000~\micron\ luminosity of the cold dust model component.  We leave the discussion of the stellar mass to Section \ref{ssec:mh2}.  The axis ratio ($q$) and inclination angle ($i$) are discussed in Section \ref{ssec:tfr}.

We compile the morphology of the quasar host galaxies derived mainly from high-resolution Hubble Space Telescope (HST) optical/near-IR images analyzed by \cite{Kim2008b, Kim2017ApJS} and Y. Zhao et al. (in preparation).  Largely based on their analysis, we visually classify the host galaxy morphology as disk, elliptical, or merger.  Considering the challenges of decomposing the bright active nucleus from the host galaxy, we do not distinguish between disk galaxies and ellipticals and simply refer to them as non-mergers.  Fortunately, it is relatively straightforward to recognize ongoing mergers and tidally interacting systems merely from visual inspection.\footnote{Some widely separated mergers (e.g., PG~0007+106 and PG~0844+340) were not recognized as mergers in the referenced papers.}  We are motivated to identify such cases to ascertain whether the molecular gas properties of the host galaxies are impacted by external dynamical perturbations.  For these reasons we fairly liberally label a host as a ``merger'' if it contains obvious tidal features or companion galaxies, whether or not it has been recognized previously as such in the literature.  None of our main conclusions is affected by this choice.  As detailed in Table~\ref{tab:phy}, a minority of cases cannot be classified because of the dominance of the active nucleus.

\subsection{Upper Limits on Outflow Flux} \label{ssec:of_flux}

We search for high-velocity line emission that might be attributable to outflows by inspecting the continuum-subtracted data cube, which was generated by fitting the {\it uv}\ data with a constant value using the task \texttt{uvcontsub}.  Given the very wide ($>5000\,\mathrm{km\,s^{-1}}$) velocity band width, any line emission from the putative outflow should not be affected by the continuum subtraction, and the cleaned aperture diameter of $>20\arcsec$ corresponds to a physical diameter of $\gtrsim 20$~kpc.  We fail to detect any significant ($>5\,\sigma$) high-velocity emission that spans several channels coherently, in any of the quasars.  We scrutinize the high-velocity channels that are safely free from contamination by emission from the main galaxy disk (e.g., beyond 200--400~$\mathrm{km\,s^{-1}}$).

\begin{figure*}
\begin{center}
\includegraphics[width=0.6\textwidth]{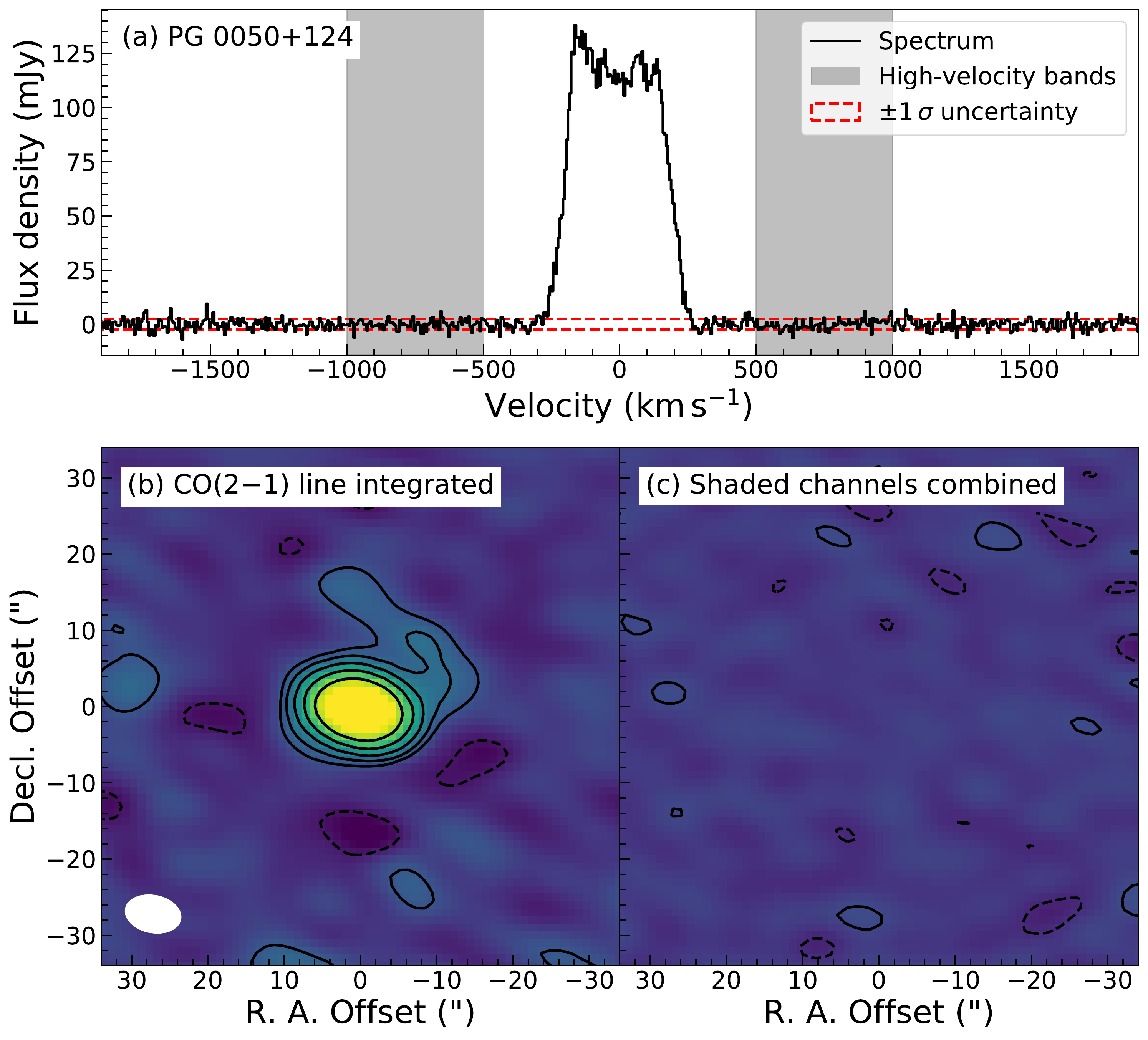}
\caption{(a) CO(2--1) spectrum extracted from a circular aperture with diameter of 10\arcsec\ and centered on the line.  The shaded channels are used to search for evidence of outflows.  (b) Intensity map of the channels including CO(2--1) line emission; the synthesized beam is given by the ellipse on the bottom-left corner.  (c) Intensity map combining the shaded channels in (a).  The contours indicate $-2$ (dashed), 2, 4, 6, 8, 16, and $32\,\sigma$ levels, with $\sigma$ being the rms of the source-free pixels in the map.}
\label{fig:outflow_ex}
\end{center}
\end{figure*}

Outflow emission is considered to be robustly detected in CO if it spans several hundreds $\mathrm{km\,s^{-1}}$ in high-velocity channels \citep{Cicone2014AA}.  Since we see no evidence of high-velocity CO(2--1) emission in any of the data cubes,  we place a $3\,\sigma$ upper limit on high-velocity emission from the intensity map generated by combining the velcity channels between $-1000$ to $-500\, \mathrm{km\,s^{-1}}$ and $+500$ to $+1000\, \mathrm{km\,s^{-1}}$.  This velocity range is typically adopted by previous works, such as those of \cite{Cicone2014AA} and \cite{Lutz2020AA}.  As the maximum velocity of quasar outflows scales with AGN luminosity, we use the empirical relation of \cite{Fiore2017AA} to set an upper bound of $1000\,\mathrm{km\,s^{-1}}$ for our PG quasars.  Unknown projection effects make it difficult to decide on an appropriate lower-bound velocity that guarantees escape from the host galaxy, so we turn to previous observations for guidance.  \cite{Cicone2014AA} identify outflow emission with velocities $> 500\,\mathrm{km\,s^{-1}}$ or line wing emission with $v>300\,\mathrm{km\,s^{-1}}$ that deviates from rotation.  Figure \ref{fig:outflow_ex} shows, for illustration purposes, the integrated spectrum, CO(2--1) line intensity map, and the map combining the high-velocity channels for PG~0050+124.  As in \cite{Cicone2014AA}, the spectrum was extracted from a 10\arcsec-diameter circular aperture centered on the emission line.  We also use the same circular aperture with diameter of 10\arcsec\ to measure the flux on the high-velocity channel map, and sample the source-free areas to estimate the uncertainty.  In no case is the measured outflow flux larger than 3 times the uncertainty.  The 10\arcsec-diameter aperture always exceeds the CO disk sizes (see Section \ref{ssec:size}), which are used to estimate the mass outflow rates.  This guarantees that the upper limits are conservative.

\subsection{Sizes of CO-emitting Region} \label{ssec:size}

Following \cite{Cicone2014AA}, we estimate the size of the CO emission by fitting the visibility data.  We first split the visibilities for the core ($\pm \mathrm{\onehalf FWHM}$ around the center) of the emission line.  We then fit the visibility data with the CASA task \texttt{uvmodelfit} using a two-dimensional Gaussian model.  If the data quality is good, we allow the axis ratio and position angle of the elliptical model to be free parameters of the fit.  However, for marginal data quality the best-fit axis ratio and/or the position angle may not always be physical, and under these circumstances we fix the model to be circular (axis ratio = 1).  

\begin{figure}
\begin{center}
\includegraphics[width=0.4\textwidth]{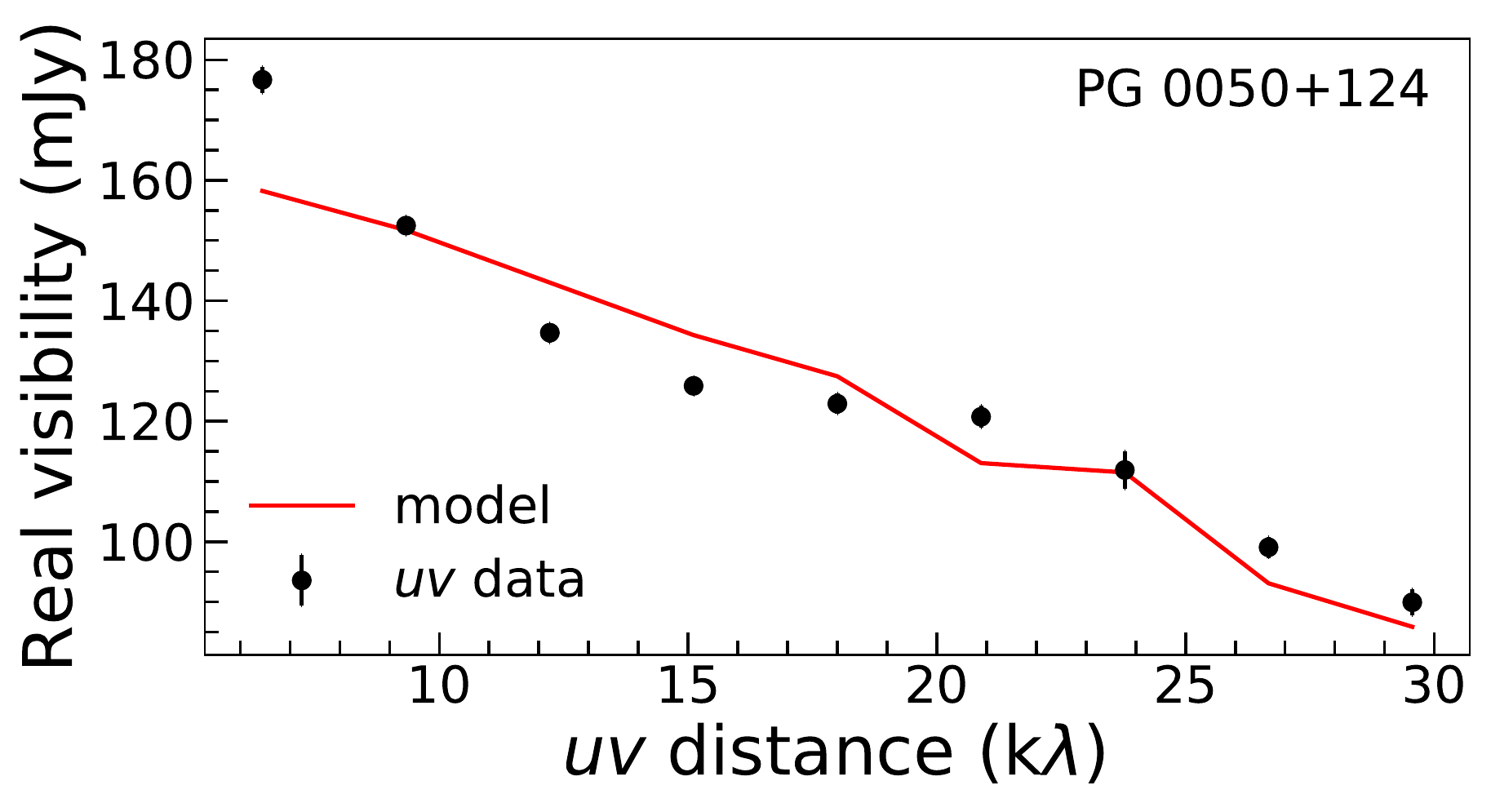}
\caption{The real part of the visibility data and the best-fit model, averaged using the \texttt{visstat} task in CASA, for illustration purposes.  We fit the full visibility data using a Gaussian model with the \texttt{uvmodelfit} task.}
\label{fig:uvfit}
\end{center}
\end{figure}

As an example, the averaged real visibility as a function of $uv$ distance of PG~0050+124 is shown in Figure~\ref{fig:uvfit}.  As the best-resolved object in our sample, the simple one-component model does not fit the data perfectly.  This is likely due to the complexity of the extended tidal arm to the northwest of the galaxy  (Figure~\ref{fig:outflow_ex}b; see below).  However, considering the purpose of our estimate, we do not consider more complicated models.  We also use the two-dimensional fitting tool of CASA to fit the intensity map of each target.  The tool successfully provides the sizes of the CO emission for less than half of the sample, but whenever measurable, the sizes from two-dimensional fitting are consistent with the results from \texttt{uvmodelfit} within the uncertainty.  The reduced $\chi^2$ reported by \texttt{uvmodelfit} is usually close to unity (Table~\ref{tab:of}).  The visibility data of PG~0923+129 and PG~1011$-$040 show similar complexity as PG~0050+124, but the size estimates from the one-component model are good enough for our purposes.

As discussed in detail in Appendix~\ref{apd:size}, the measured sizes are usually smaller than the synthesized beams, whose major axis FWHM ranges from 6\arcsec\ to 8\arcsec.  Simulating observations with CASA, we demonstrate that the size can be robustly measured when the source is larger than 1\arcsec.  None of our measured CO major axis FWHM is below this limit (Table~\ref{tab:of}).  Moreover, for six quasars\footnote{PG~0050+124, PG~0923+129, PG~1011$-$040, PG~1126$-$041, PG~1244+026, and PG~2130+099.} with high-resolution (beam size $\lesssim 1\arcsec$) ALMA observations, the CO radii constrained from the ACA data are consistent within 30\% of the half-light radii measured by J. Molina et al. (in preparation).  The only exception is PG~0050+124, whose ACA-derived size is 50\% higher.  J. Molina et al. fit a S\'{e}rsic (1968) profile to the intensity maps and found S\'{e}rsic indices $\lesssim 1$ (close to a Gaussian profile), and so our measured sizes are directly comparable.  The high-resolution CO map of PG~0050+124 reveals a compact core plus two spiral arms.  The size from the ACA $uv$ data is likely affected by the spiral arms, in particular the more extended one to the northwest.  In any event, the comparison strongly indicates that our size estimates well characterize the overall size of the CO emission.

The \texttt{uvmodelfit} task can also fit a two-dimensional disk model, but the goodness-of-fit is always similar to or slightly worse than that for the Gaussian model.  The major axis of the best-fit disk model is on average a factor of $\sim 1.6$ larger than that of the Gaussian model, while the axis ratio and position angle are similar between the two models. We prefer to adopt the sizes from the Gaussian model in order to provide more conservative estimates of the mass outflow rates (see Section \ref{ssec:outflow}).

\section{Results} 
\label{sec:res}

\subsection{Molecular Gas Mass}
\label{ssec:mh2}

\begin{figure}
\begin{center}
\includegraphics[height=0.3\textheight]{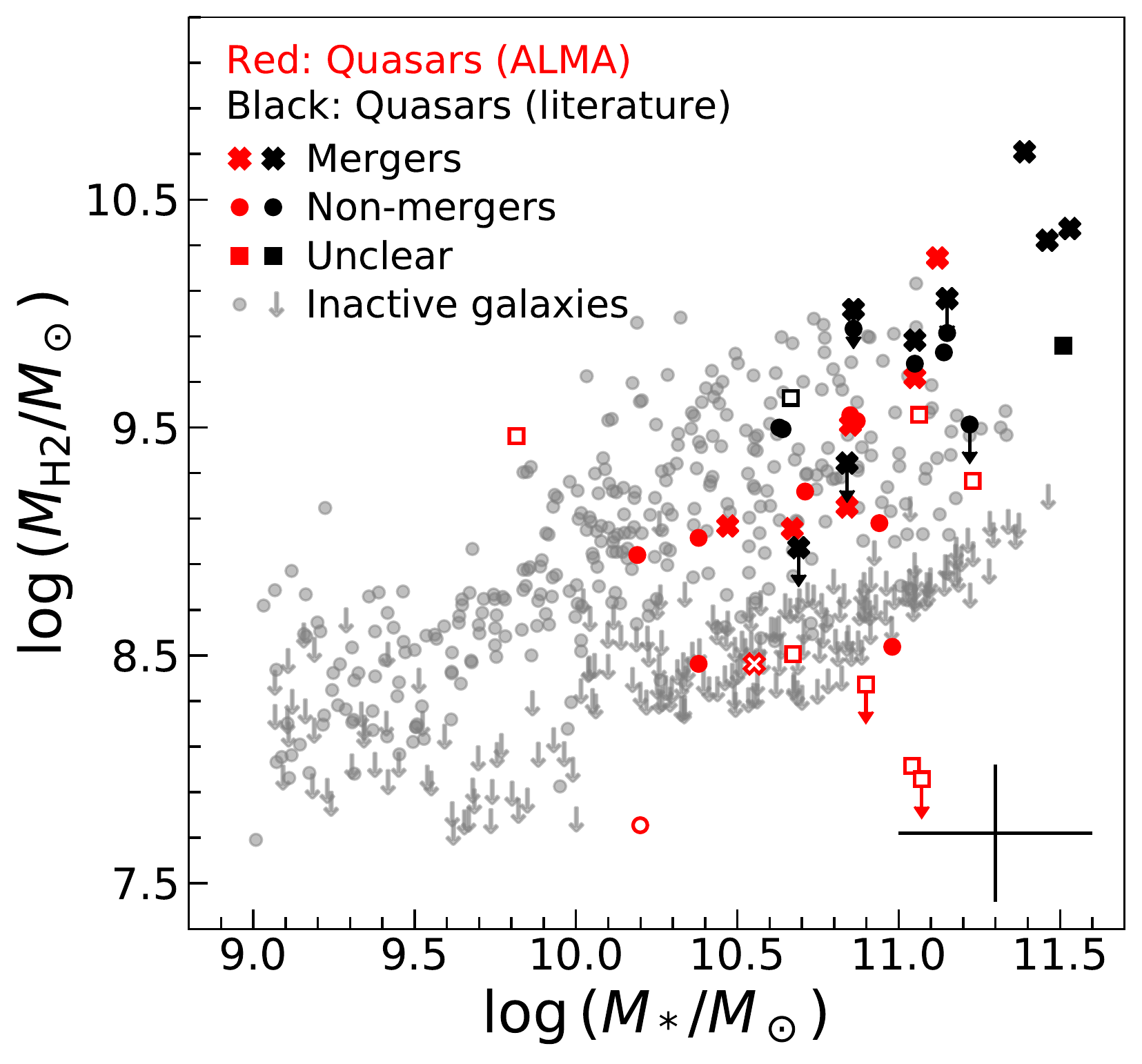}
\caption{
The molecular gas masses of the quasars show a wide range that is generally consistent with that of inactive galaxies in the xCOLD GASS sample.  Our new measurements reveal some quasars residing well in the regime of gas-poor galaxies.  The molecular gas mass is calculated with $\aco=3.1$ \uaco, as recommended by \cite{Sandstrom2013ApJ}.  The quasars in our ALMA sample are in red, while those from the literature are in black.  The morphology of the quasar host galaxies is classified into three types based on high-resolution \hst\ images: mergers (crosses), non-mergers (circles), and unclear (squares).  The filled symbols denote the quasars with directly measured stellar masses, while the open symbols denote those with indirect stellar masses.  The inactive galaxies are shown with filled grey circles, with grey arrows denoting CO upper limits.  Typical uncertainties are plotted on the lower-right corner.
}
\label{fig:mh2}
\end{center}
\end{figure}

The molecular gas content of galaxies varies with stellar mass \citep{Saintonge2016}, and hence any discussion of the gas content of quasar host galaxies must consider how to estimate their stellar mass ($M_*$).  This is non-trivial, in view of the severe contamination of the starlight by the bright and sometimes overwhelming nonstellar nucleus \citep{Kim2008a}.  Direct estimates of $M_*$ are available for 30 of the quasar host galaxies, based on decomposition of high-resolution near-IR images by \cite{Zhang2016ApJ}.\footnote{The stellar masses of PG~0923+129, PG~0934+013, PG~1011$-$040, PG~1244+026, and PG~1448+273 are supplemented by new estimates based on $B$-band and $I$-band \hst\ photometry (PI: L.~C.~Ho) analyzed by Y. Zhao et al. (in preparation).}  For the remaining quasars, \cite{Shangguan2018ApJ} obtained lower limits to the total $M_*$ by estimating the contribution from the bulge component alone using the empirical correlation between bulge stellar mass and BH mass \citep{Kormendy2013ARAA}.  Here we adopt a different, improved strategy, one that obviates the uncertainty introduced by the poorly determined bulge-to-disk ratio of the host.   We predict the total stellar mass from the observed $M_{\rm BH}-M_*$ relation of early-type galaxies, as recently calibrated by \cite{Greene2020}:

\begin{equation}\label{eq:ms}
\log \left(\frac{M_{\rm BH}}{M_\odot}\right) = (7.89\pm0.09) + (1.33\pm0.12)\log \left(\frac{M_*}{3\times10^{10}\,M_\odot}\right),
\end{equation}

\noindent
which has an intrinsic scatter of 0.65 dex.  Using the subsample with directly measured stellar masses as a cross-check, we find that Equation~\ref{eq:ms} underpredicts the direct measurements by 0.2$\pm$0.4~dex, which is consistent with the intrinsic scatter.  Figure \ref{fig:mh2} shows the variation of $M_{\rm H_2}$ as function of $M_*$. The molecular gas masses of the quasar host galaxies span a wide range, but are in general consistent with those of normal galaxies of similar stellar mass.  The CO-detected quasars have $M_{\rm H_2} = 10^{7.76} - 10^{10.98}\,M_\odot$, with a mean value of $10^{9.20\pm0.13}\,M_\odot$ after accounting for the upper limits using the Kaplan-Meier estimator\footnote{Implemented as the \texttt{kmestimate} task in \texttt{IRAF.ASURV}.} \citep{Feigelson1985ApJ, Lavalley1992ASPC}.  The high sensitivity of ALMA allows us to detect molecular gas masses or provide stringent upper limits thereof in the regime of gas-poor galaxies ($\mmol \lesssim 10^{8.5}\,M_\odot$ or $\mmol/M_* \lesssim 0.01$; \citealt{Saintonge2016}) for $\sim 35\%$ for our ALMA sample.  Our results qualitatively confirm the conclusions of \cite{Shangguan2018ApJ}, who estimated total gas masses from cold dust emission for the 87 PG quasars with $z < 0.5$.  They found a somewhat higher fraction of gas-rich systems than we, likely because their sample includes more higher redshift systems.

\subsection{CO Tully-Fisher Relation of Quasars}
\label{ssec:tfr}

\begin{figure}
\begin{center}
\includegraphics[height=0.3\textheight]{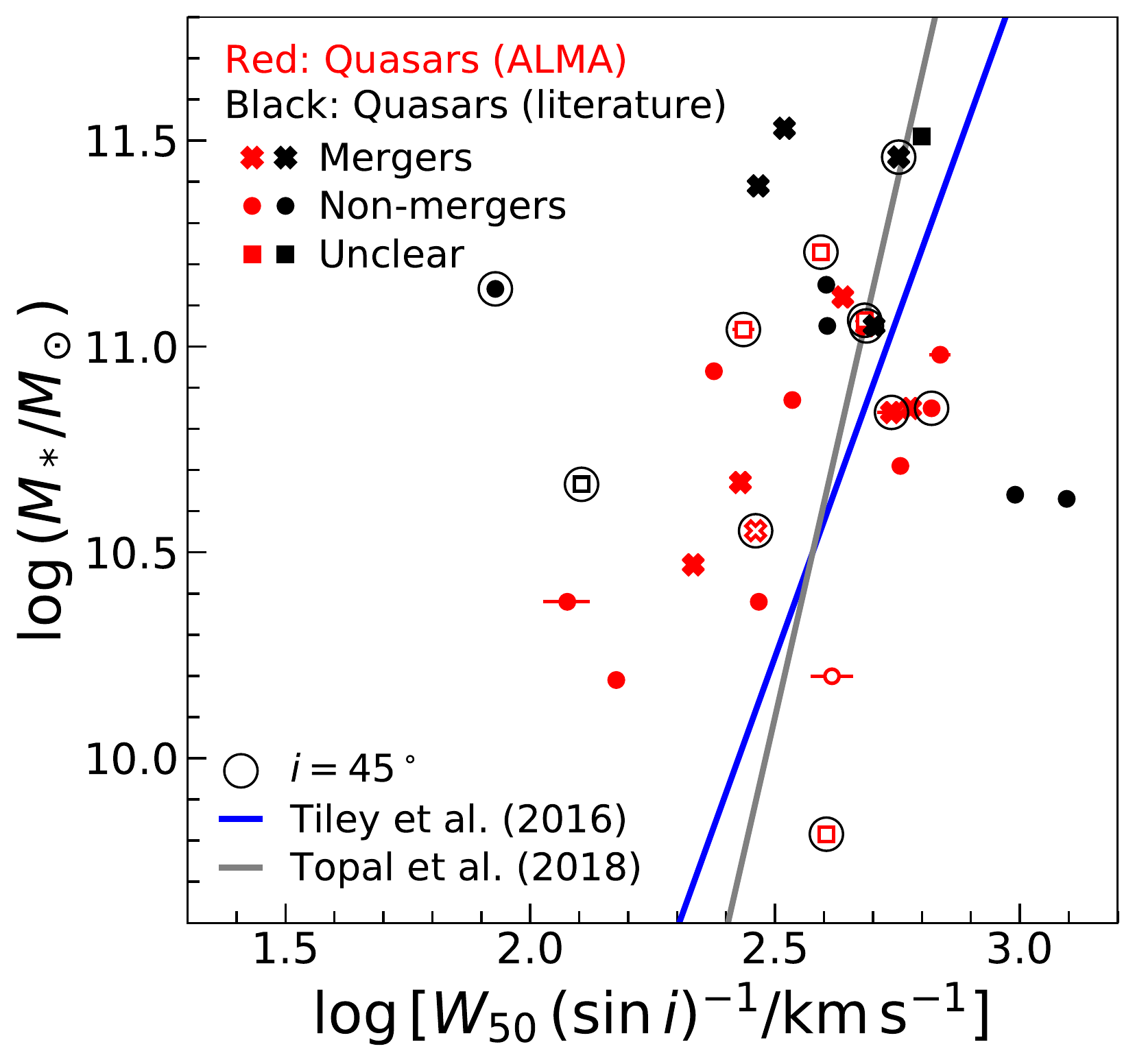}
\caption{The Tully-Fisher relation of the PG quasars are compared with the relation of inactive galaxies.  Symbols and colors follow Figure ~\ref{fig:mh2}, with additions as follows.  Objects enclosed with a large black circle do not have measurement of inclination angle, and we assume $i=45\degree$.  The blue line is the Tully-Fisher relation of galaxies in the local universe from the COLD GASS sample \citep{Tiley2016MNRAS}; the grey line is the relation of galaxies at $z = 0.05-0.3$ from \cite{Topal2018MNRAS}, which better match the redshift range of our quasars.  The horizontal error bars only consider the uncertainty of $W_{50}$, which is smaller than the symbol size for most of the cases.  We do not consider the uncertainty of the inclination angle, which is likely more important (see the main text).  To avoid complication, we do not show the uncertainty associated with the stellar mass.
}
\label{fig:tfr}
\end{center}
\end{figure}

The nature of quasar host galaxies can be constrained not only by the amount but also the dynamical state of their molecular gas.  Is the gas virialized, or is it being blown out of the galaxy by strong quasar-mode feedback?  While our ALMA observations lack the spatial resolution to map the velocity field of the gas, we can still derive some rudimentary kinematic constraints from the integrated line width of the CO emission.  By analogy with the more familiar \cite{Tully1977AA} relation based on \hi\ 21~cm emission, we can define a CO Tully-Fisher relation for normal galaxies \citep{Dickey1992, Sofue1992}, which can be extended further to the host galaxies of AGNs and quasars \citep{Ho2007}.  To correct the observed CO line width for projection effect, we assume that the gas is coplanar with the stars and estimate the inclination angle $i$ from the prescription of \cite{Hubble1926ApJ},
 
\begin{equation}
\cos^2 i  = \frac{q^2 - q^2_0}{1 - q^2_0}, 
\end{equation}

\noindent
where $q$ is the ratio of the semi-minor to semi-major axis of the stars, which we obtain from the \galfit\ \citep{Peng2002AJ,Peng2010AJ} model of the host galaxy (\citealt{Kim2017ApJS}; Y. Zhao et al. in preparation).  The intrinsic thickness of the disk is assumed to be $q_0 = 0.2$ for late-type galaxies, but the results are not significantly different if we adopt $q_0=0.34$ for early-type galaxies \citep{Tiley2016MNRAS}. For models with more than one component, we use $q$ of the disk component.   We assume $i=45\degree$ if no suitable images of the host are available.  Despite the large scatter, it is interesting that PG quasars follow essentially the same CO Tully--Fisher relation of inactive galaxies (Figure \ref{fig:tfr}).  Three objects (PG~0838+770, PG~1211+143, PG~1415+451) stand out as strong outliers with $M_*\gtrsim10^{10.4}\,M_\odot$ and $W_{50}(\sin\,i)^{-1} \lesssim 125\,\mathrm{km\,s^{-1}}$, most likely because they are almost face-on and suffer large uncertainties.  We only have an inclination angle estimate for PG~1211+143, which, indeed, is close to face-on.  Two objects have abnormally large deprojected line widths [$W_{50}(\sin\,i)^{-1} \gtrsim 1000\,\mathrm{km\,s^{-1}}$].  The full width at zero intensity of PG~0804+761 was reported as $881\,\kms$ \citep{Scoville2003ApJ}, but the measured line flux significance is only $4\,\sigma$ and thus the line width may be overestimated.  PG~1351+640 has a typical line width ($W_{50} = 260\,\mathrm{km\,s^{-1}}$), but the nearly face-on orientation of the host galaxy results in a large and uncertain inclination correction.

\subsection{Molecular Gas and AGN Fueling}
\label{ssec:lcoagn}

\begin{figure*}
\begin{center}
\includegraphics[height=0.3\textheight]{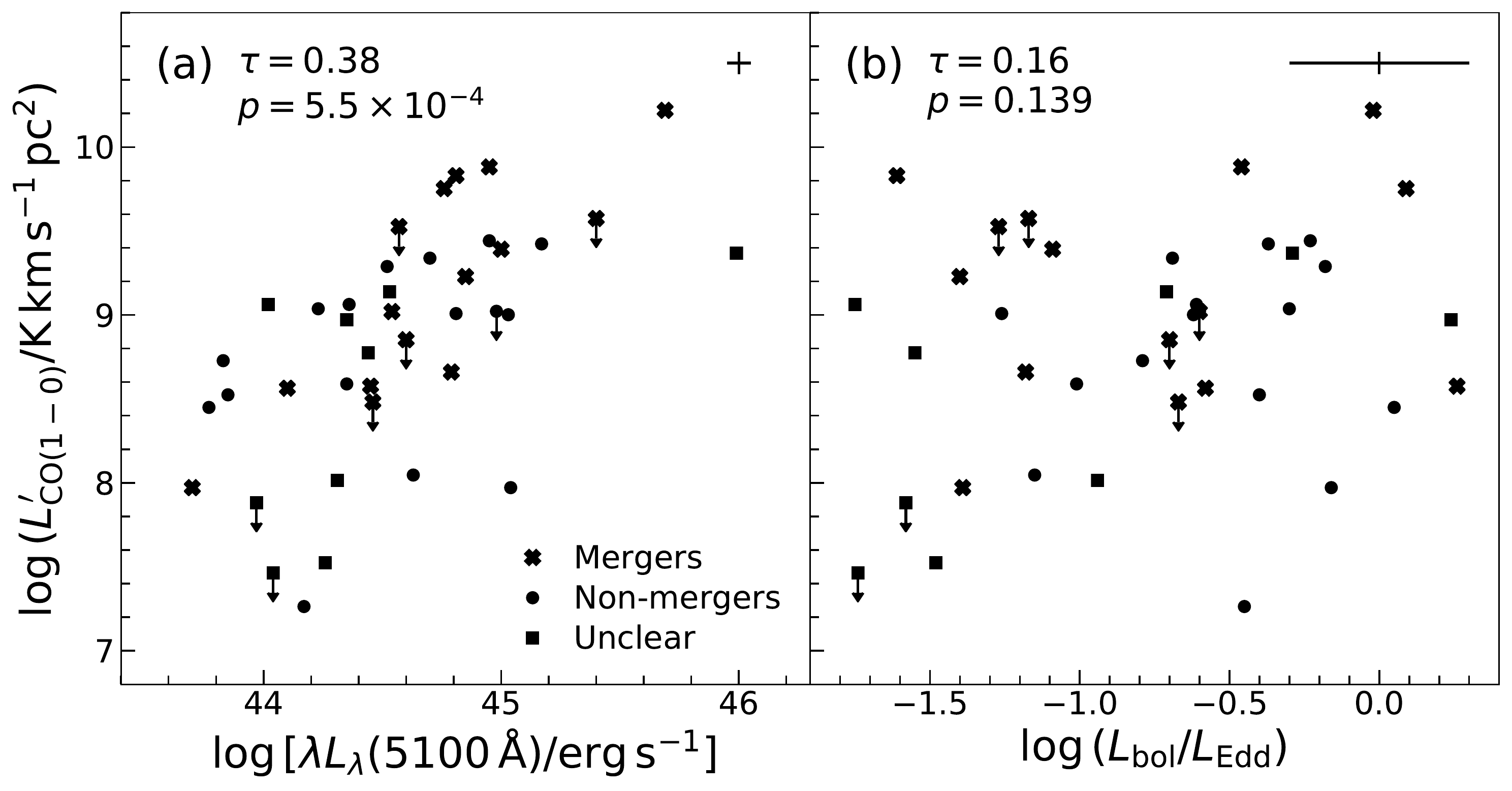}
\caption{The CO luminosity (\lco) correlates significantly with (a) AGN luminosity \lagn\ but not with (b) the Eddington ratio.  The generalized Kendall's correlation coefficient ($\tau$) and the corresponding $p$-value, accounting for the upper limits, are given on the upper-left corner of each panel.   The morphologies of the quasar host galaxies are classified into mergers (crosses), non-mergers (circles), and unclear (squares).  The typical uncertainties are plotted on the upper-right corner.  For clarity, the uncertainty of the vertical axis has been increased by a factor of 3.
}
\label{fig:co51}
\end{center}
\end{figure*}

Figure \ref{fig:co51} studies the variation of \lco\ with the 5100~\AA\ continuum luminosity as well as the Eddington ratio of the AGN.  We assume that the bolometric luminosity is given by $L_\mathrm{bol} = 10\,\lambda L_\lambda(5100\,\mathrm{\AA})$ \citep{McLure2004MNRAS,Richards2006ApJS}, and the Eddington luminosity is defined as $L_\mathrm{Edd}=1.26 \times 10^{38}\, (M_\mathrm{BH}/M_\odot)\, \mathrm{erg\,s^{-1}}$.  We use the generalized Kendall's $\tau$ calculated with the \texttt{cenken} function from the NADA package of \texttt{R} to quantitatively test the significance of correlation of different quantities including censored data.  Throughout the paper, we consider a correlation significant if the $p$-value of the null hypothesis that there is no correlation between the two quantities is $<0.01$, and we consider the correlation moderately significant if $p \approx 0.01-0.05$.  We find that the \lco--\lagn\ correlation is significant with $\tau=0.38$ and $p=5.5\times 10^{-4}$.  The correlation of the merger subsample alone ($\tau=0.51$ and $p=6.8\times 10^{-3}$) is more significant than that of the non-merger subsample ($\tau=0.32$ and $p=0.03$).  We checked that distance is not the driving factor in any of the luminosity correlations.  Restricting the sample with $z < 0.15$ to mitigate the possible redshift dependences, the \lco--\lagn\ correlation remains moderately significant ($\tau=0.32$ and $p=0.013$), although the sample size is small.  Using optical extinction to indirectly infer the molecular gas mass \citep{Yesuf2019ApJ} of a large sample of AGNs, M.-Y. Zhuang et al. (in preparation) also find a significant correlation between the molecular gas and AGN luminosity.  By comparison, \lco\ shows no clear trend with the Eddington ratio.  These results are consistent with \cite{Husemann2017MNRAS}, who interpreted the \lco--\lagn\ correlation they found as a link between the BH accretion rate and the gas reservoir (see below for more discussion).  

We fit the \lco--\lagn\ relation with \linmix\ \citep{Kelly2007ApJ}, including the censored data and setting \lco\ as the dependent variable.  Assuming a uniform uncertainty of 0.05 dex for \lagn\ \citep{Vestergaard2006ApJ} and a conservative uncertainty of 0.1 dex for \lco, we find $L^{\prime}_\mathrm{CO} \propto \lambda L_\lambda(5100\,\mathrm{\AA})^{0.76\pm0.20}$ with an intrinsic scatter of $0.36^{+0.12}_{-0.08}$ dex.  This agrees well with \cite{Xia2012ApJ}, who found $\lco \propto \lagn^{0.71}$, in their study of ultraluminous IR quasars combined with nearby and high-redshift quasars.

\begin{figure}
\begin{center}
\includegraphics[height=0.3\textheight]{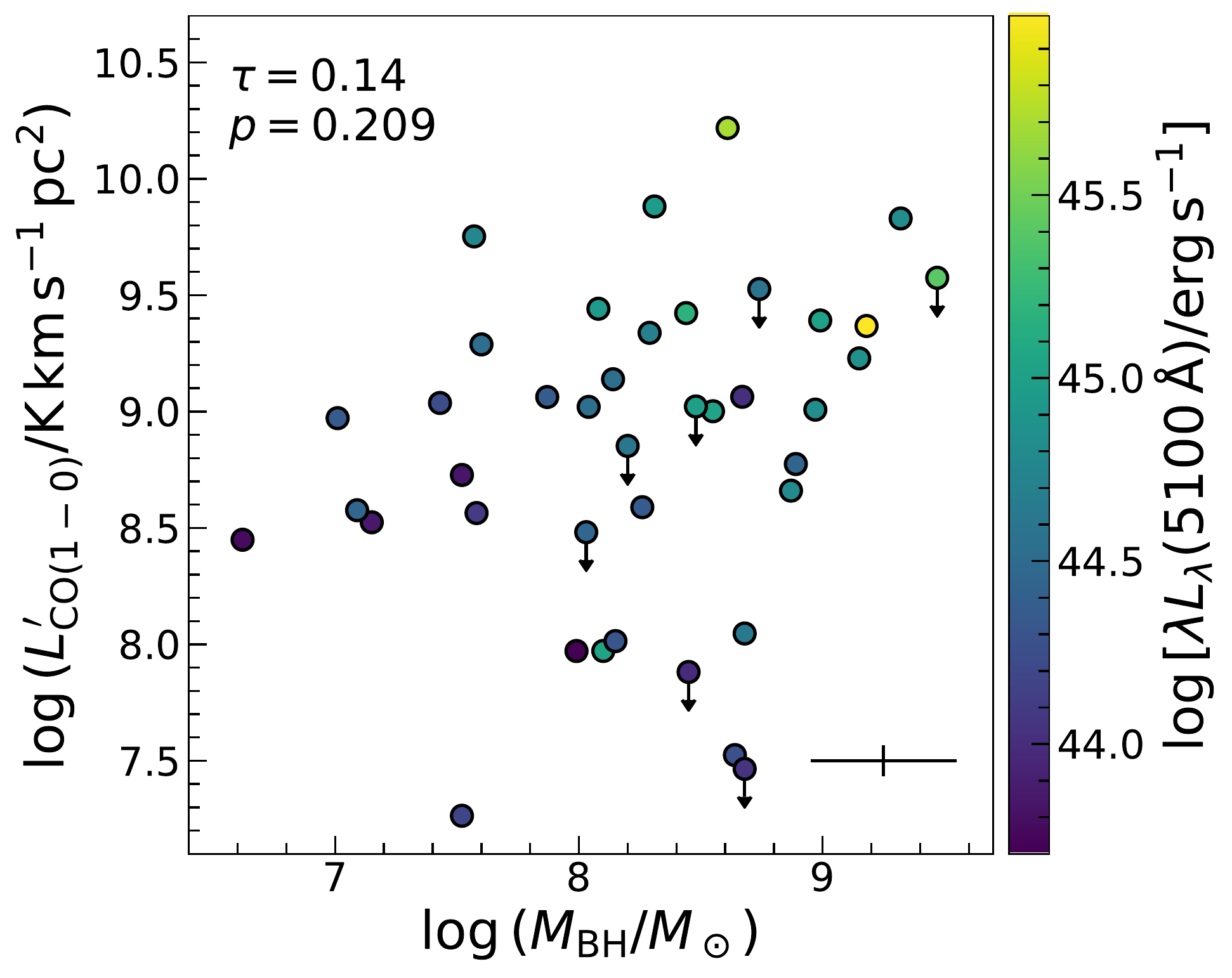}
\caption{The AGN luminosity, \lagn, increases as both BH mass and \lco\ increase, while BH mass and \lco\ themselves are not correlated significantly.  The generalized Kendall's correlation coefficient ($\tau$) and the corresponding $p$-value, accounting for the upper limits, are shown on the upper-left corner.  Typical uncertainties are plotted on the lower-right corner.  For clarity, the uncertainty of the vertical axis has been increased by a factor of 3.
}
\label{fig:mbh}
\end{center}
\end{figure}

On the one hand, the relation between \lco\ and \lagn\ suggests a close connection between BH accretion and cold gas supply, especially gas in the central sub-kpc scale \citep{DiamondStanic2012ApJ,Xia2012ApJ,Esquej2014ApJ,Izumi2016ApJ, Husemann2017MNRAS,Lutz2018AA}.  On the other hand, the relation may be secondary, reflecting the common dependence of \lco\ and \lagn\ on galaxy stellar mass, and hence BH mass.  Figure~\ref{fig:mbh} shows, however, that \lco\ does not correlate significantly with BH mass, whereas the clear gradient from the lower-left to the upper-right corner of the diagram suggests that \lco\ and BH mass affect \lagn\ independently. 
 
\begin{figure}
\begin{center}
\includegraphics[width=0.4\textwidth]{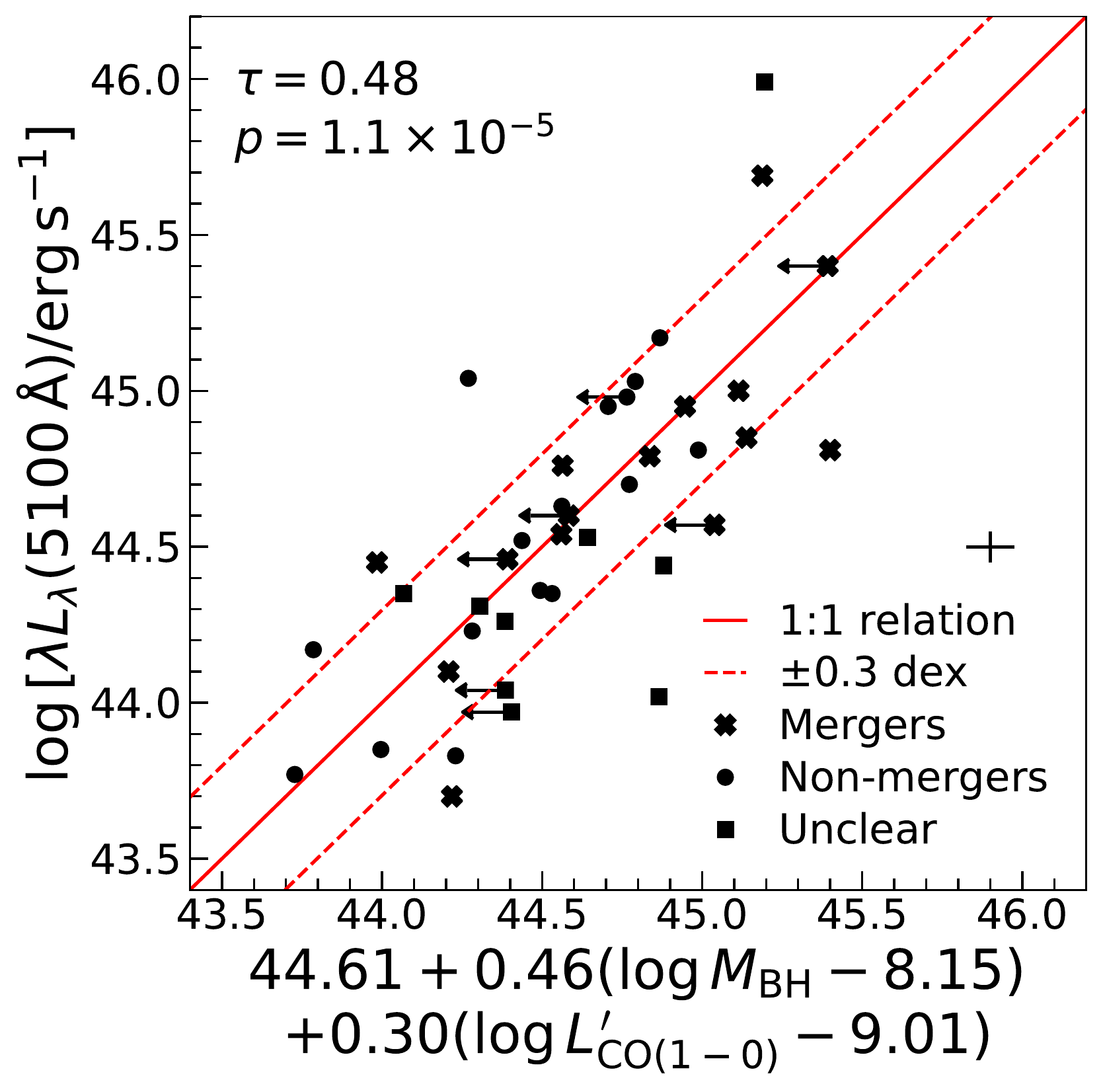}
\caption{The projected relation of the best-fit plane of \mbh, \lco, and AGN continuum luminosity at 5100~\AA, computed using  \texttt{LTS\_PLANEFIT} \citep{Cappellari2013MNRAS}.  The intrinsic scatter is 0.3 dex.  Objects with \lco\ upper limits are plotted in the figure but not included in the fit.  The generalized Kendall's correlation coefficient ($\tau$) and the corresponding $p$-value, including the upper limits on the horizontal axis, are displayed on the upper-left corner.  The typical uncertainties are plotted on the lower-right corner.  The morphologies of the quasar host galaxies are classified into mergers (crosses), non-mergers (circles), and unclear (squares).}
\label{fig:mbhproj}
\end{center}
\end{figure}

We fit the three quantities with a plane, using the widely used code \texttt{LTS\_PLANEFIT} \citep{Cappellari2013MNRAS}, which incorporates a least trimmed squares technique to iteratively clip out outliers.  We choose the clip threshold to be $3\,\sigma$, so that all of the data points are used to obtain the best-fit relation.  The code does not allow us to include the objects with \lco\ upper limits, but, as Figure~\ref{fig:mbhproj} shows, these objects (x-axis upper limits) are unlikely to affect the results significantly.  The best-fit plane is given by 

\begin{eqnarray}\label{eq:plane}
&& \log\,\lagn = (44.61 \pm 0.06) + (0.46 \pm 0.09) \\ \nonumber
&& \times (\log\,\mbh - 8.15) + (0.30 \pm 0.09) \times (\log\,\lco - 9.01),
\end{eqnarray}

\smallskip
\noindent
where the units of \lagn, \mbh, and \lco\ are \ergs, $M_\odot$, and $\mathrm{(K\,km\,s^{-1}\,pc^{2})^{-1}}$, respectively, and the intrinsic scatter is 0.3 dex.  The objects classified as mergers or non-mergers do not show distinctive behavior.  The correlation between \lagn\ and the projected horizontal axis is more significant ($\tau = 0.48$, $p=1.1\times10^{-5}$) than the relation between \lagn\ and \mbh\ (not shown; $\tau = 0.41$, $p=2.3\times10^{-4}$) or \lagn\ and \lco\ (Figure~\ref{fig:co51}a; $\tau = 0.38$, $p=5.5\times10^{-4}$).  The partial correlation of \lagn\ and the projected horizontal axis is still significant ($\tau=0.29$, $p=8\times10^{-3}$) after their mutual dependences on the luminosity distance are removed.  This suggests that the correlation among the three quantities is physical.  We emphasize that the BH mass calculated with the single-epoch method is $\mbh \propto \lagn^{0.533}$ \citep{Ho2015ApJ}, and so the dependence of the BH mass in Equation~\ref{eq:plane} is not trivially born from the estimate of the \mbh.

Why does the AGN luminosity depend on both BH mass and molecular gas mass?  We do not have a definitive, quantitative answer, but we offer some speculations.  At the most rudimentary level, AGNs, of course, need to be powered by accretion of material.  For AGNs powerful enough to be deemed quasars, most of the material must derive from a suitably plentiful reservoir of cold gas, which naturally takes the form of a circumnuclear disk (e.g., \citealt{Kawakatu2008ApJ, Husemann2017MNRAS}).  Residual debris from local stellar mass loss or the occasional tidal disruption of a star can sustain the fuel requirements of low-luminosity AGNs ($L_\mathrm{bol}/L_\mathrm{Edd} \lesssim 0.01$; \citealt{Ho2008}), but not quasars.  Still, the hot plasma in the central regions of galactic bulges will contribute to the fueling budget as it undergoes \cite{Bondi1952} accretion \citep{Ho2009}, at a rate that depends on BH mass and gas temperature as $\dot{M}_\mathrm{B}\propto M_\mathrm{BH}^2 T_\mathrm{gas}^{-3/2}$ (e.g., \citealt{Inayoshi2019MNRAS,Inayoshi2020}).  If the gas is close to virialized, $T_\mathrm{gas} \propto M_\mathrm{BH}$, and $\dot{M}_\mathrm{B}\propto M_\mathrm{BH}^{0.5}$.  Interestingly, this is consistent with our fitting result: $\lagn \propto M_\mathrm{BH}^{0.46\pm0.09}$.

\subsection{Relation between AGN Luminosity and Infrared Luminosity is Driven by the Molecular Gas}
\label{sec:liragn}

\begin{figure*}
\begin{center}
\includegraphics[height=0.3\textheight]{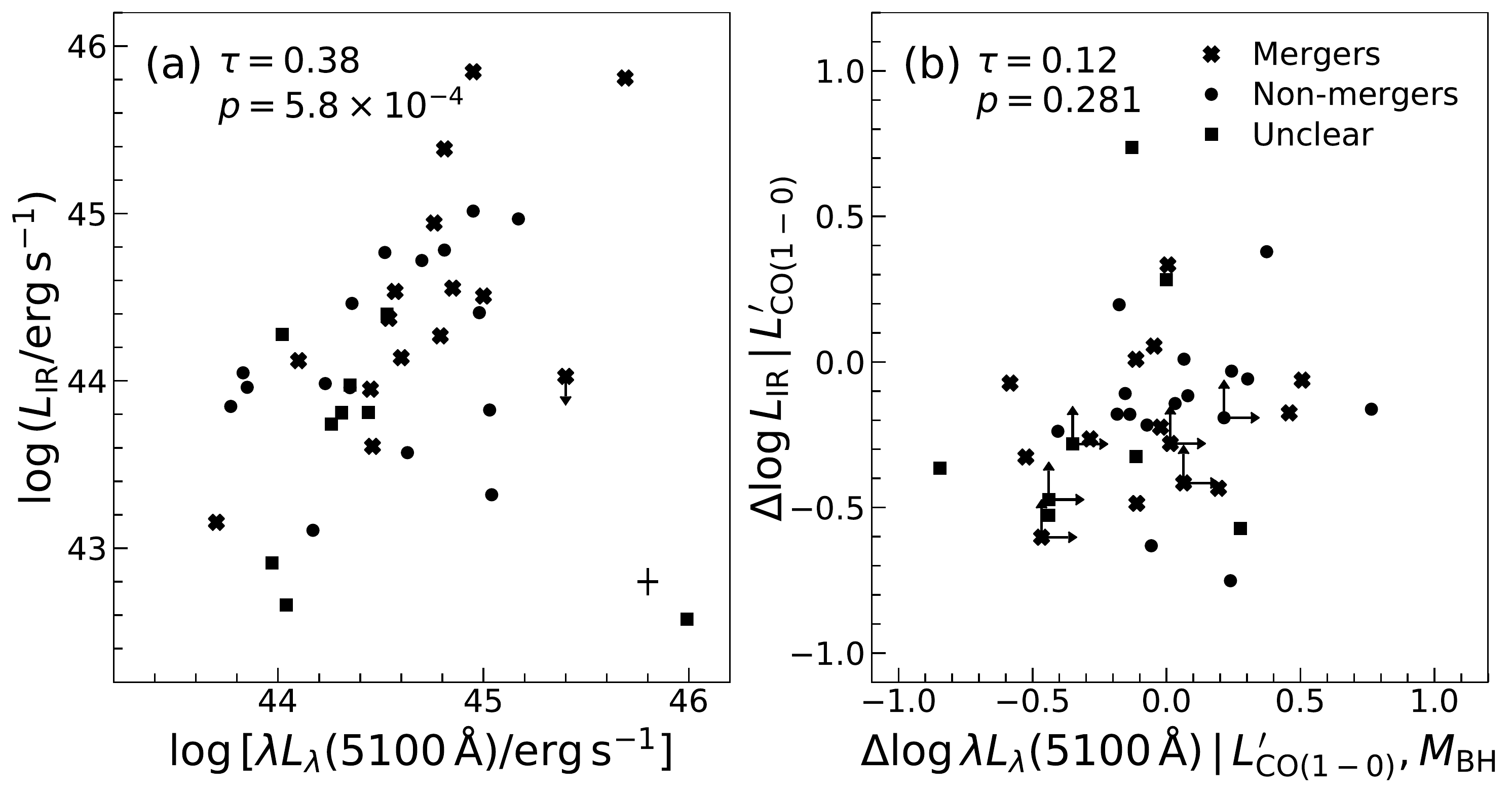}
\caption{(a) A significant correlation is found between the IR luminosity \lir\ and the AGN luminosity \lagn.  The generalized Kendall's correlation coefficient ($\tau$) and the corresponding $p$-value, including the upper limits on the horizontal axis, are displayed on the upper-left corner.  The morphologies of the host galaxies are classified into mergers (crosses), non-mergers (circles), and unclear (squares).  The quasars in the merger host galaxies show a more significant correlation than the rest of the sample.  The typical uncertainties are plotted on the lower-right corner.  For clarity, the uncertainty of the vertical axis has been increased by a factor of 3.  (b) We study the partial correlation of \lir\ and \lagn\ by removing the dependence of \lir\ on \lco\ (Equation 4 of \citealt{Shangguan2020ApJS}) and the dependence of \lagn\ on \lco\ and \mbh\ (Equation \ref{eq:plane}).  There is no significant partial correlation for the entire sample or for the individual subsamples.  The measurement uncertainties of \lco, \lagn, and \lir\ are presented in Figure~\ref{fig:co51} and panel (a), while the scatter of the data points is mainly due to the intrinsic scatter of the \lir--\lco\ and \lagn--\lco--\mbh\ relations.  PG~1226+123, the extreme outlier in the lower-right corner of panel (a), has a very uncertain \lir\ from spectral energy distribution decomposition; it is excluded from panel (b)  and from the correlation tests.}
\label{fig:ir51}
\end{center}
\end{figure*}

Many studies have discussed the correlation between AGN emission (usually measured in the X-rays or ultraviolet/optical) and host galaxy star formation (e.g., \citealt{Bonfield2011MNRAS, DiamondStanic2012ApJ,Xia2012ApJ,Xu2015ApJ,Dai2018MNRAS,Lutz2018AA, Grimmett2020arXiv, Zhuang2020}), pointing to a common link between star formation and BH accretion on the one hand and between star formation and cold gas content of the host galaxy on the other.  We, too, find that the AGN luminosity significantly correlates with the IR luminosity of the host (Figure~\ref{fig:ir51}a).\footnote{For completeness, fitting the \lir--\lagn\ relation with \texttt{Linmix} gives $\log\,\lir = 1.06\left(^{+0.22}_{-0.21}\right) \log\,\lagn - 2.95\left(^{+9.58}_{-9.34}\right)$.  PG~1226+023 and PG~1545+210 have large uncertainties on \lir\ and are excluded from the fit.}  Ambiguity exists, however, as to the interpretation of this result.  What heats the dust?  Does the quasar influence the dust on galactic scales?  This was suggested by \cite{Shangguan2018ApJ}, whose analysis of the global IR spectral energy distribution found an increase of the intensity of the interstellar radiation field with increasing quasar luminosity.  Or are we witnessing the enhancement of star formation by positive AGN feedback \citep{Maiolino2017Natur}?  Or perhaps the correlation merely trivially reflects the mutual dependence of AGN and IR luminosity on a common third variable, such as gas content.

We know that \lir\ couples strongly with \lco\ (\citealt{Shangguan2020ApJS}, their Equation 4),\footnote{The Kendall's $\tau$ and the $p$-value are 0.69 and $5.9\times 10^{-10}$, respectively.} and \lagn\ is tightly correlated with both \lco\ and \mbh\ (Figure~\ref{fig:mbhproj}).  The intrinsic scatter of both relations is only $\sim 0.3$ dex.  It is important to remove the common dependence of \lir\ and \lagn\ on molecular gas (as traced by \lco) in order to assess any possible additional influence from BH accretion.  We study the partial correlation of \lir\ and \lagn\ (Figure~\ref{fig:ir51}b) by removing the dependence of \lir\ on \lco\ (Equation~4 of \citealt{Shangguan2020ApJS}) and the joint dependence of \lagn\ on \lco\ and \mbh\ (Equation~\ref{eq:plane}).\footnote{The results do not depend on the relation between \lagn\ and \mbh.}  After taking these effects into consideration, we find that \lir\ and \lagn\ are no longer correlated.  This strongly suggests that the overall \lir--\lagn\ relation is largely driven by the mutual dependence of IR luminosity and AGN luminosity on molecular gas, which fuels both star formation and BH accretion.  It also provides a qualitative explanation for the connection between stellar mass and both the SFR and AGN luminosity (e.g., \citealt{Xu2015ApJ,Yang2017ApJ,Suh2019ApJ,Stemo2020ApJ,Ni2020}), since the molecular gas mass scales with the stellar mass of star-forming galaxies.  There is no evidence that BH accretion heats the dust on large scales, nor does AGN feedback suppress or enhance galactic star formation.  This is consistent with \cite{Xie2020}, who recently found that the SFRs of quasar host galaxies based on the far-IR continuum agree well with SFRs robustly derived from the mid-IR neon emission lines \citep{Zhuang2019ApJ}.  One caveat, however, is that revealing a statistically significant partial correlation may require a sample much larger than that considered here (M.-Y. Zhuang et al. in preparation).

\subsection{Upper Limits on Molecular Gas Outflows} \label{ssec:outflow}

Assuming that the clouds in an outflow uniformly fill a spherical or (multi-)conical volume \citep{Maiolino2012MNRAS,Cicone2014AA,Fiore2017AA}, the mass outflow rate is

\begin{equation}\label{eq:of}
\dot{M}_\mathrm{H_2, out} = 3 v \frac{M_\mathrm{H_2, out}}{R_\mathrm{out}},
\end{equation}

\noindent
with $v$ the velocity, $M_\mathrm{H_2, out}$ the molecular hydrogen mass, and $R_\mathrm{out}$ the radius of the outflow.  While the assumption of the outflow history systematically affects the estimate of the outflow rate, Equation (\ref{eq:of}) gives a factor of 3 larger outflow rate than that derived from assuming a constant outflow history \citep{Lutz2020AA}.  It thus represents a conservative 
upper limit.

Adopting the maximum velocity ($1000\,\mathrm{km\,s^{-1}}$) used to estimate the upper limits on outflow flux (Section \ref{ssec:of_flux}), the uncertainty of the mass outflow rate follows from 

\begin{equation}\label{eq:ofd1}
\Delta \dot{M}_\mathrm{H_2, out} = 3 v \sqrt{\left(\frac{\Delta M_\mathrm{H_2, out}}{R_\mathrm{out}}\right)^2 + 
\left(\frac{M_\mathrm{H_2, out}\Delta R_\mathrm{out}}{R_\mathrm{out}^2}\right)^2},
\end{equation}
 
\noindent
where $\Delta M_\mathrm{H_2, out}$ is the uncertainty of the molecular gas mass and $\Delta R_\mathrm{out}$ is the uncertainty of the radius of the outflow.  Since the outflow is not detected, we restrict ourselves to consider only the outflow within the size of the molecular disk.  With $M_\mathrm{H_2, out} \lesssim 3\, \Delta M_\mathrm{H_2, out}$, $R_\mathrm{out} = R_{\scriptsize \coline}$, and $\Delta R_\mathrm{out} \lesssim R_{\scriptsize \coline}$, we have
 
\begin{equation}\label{eq:ofd2}
\Delta \dot{M}_\mathrm{H_2, out} \lesssim 3\sqrt{10} v \frac{\Delta M_\mathrm{H_2, out}}{R_{\scriptsize \coline}}.
\end{equation}
 
\noindent
A conservative estimate of the $3\,\sigma$ upper limit of the mass outflow rate is therefore $9\sqrt{10}v (\Delta M_\mathrm{H_2, out}/R_{\scriptsize \coline})$.  The factor $\sqrt{10}$ includes the uncertainty of the radius.

\begin{figure}
\begin{center}
\includegraphics[height=0.3\textheight]{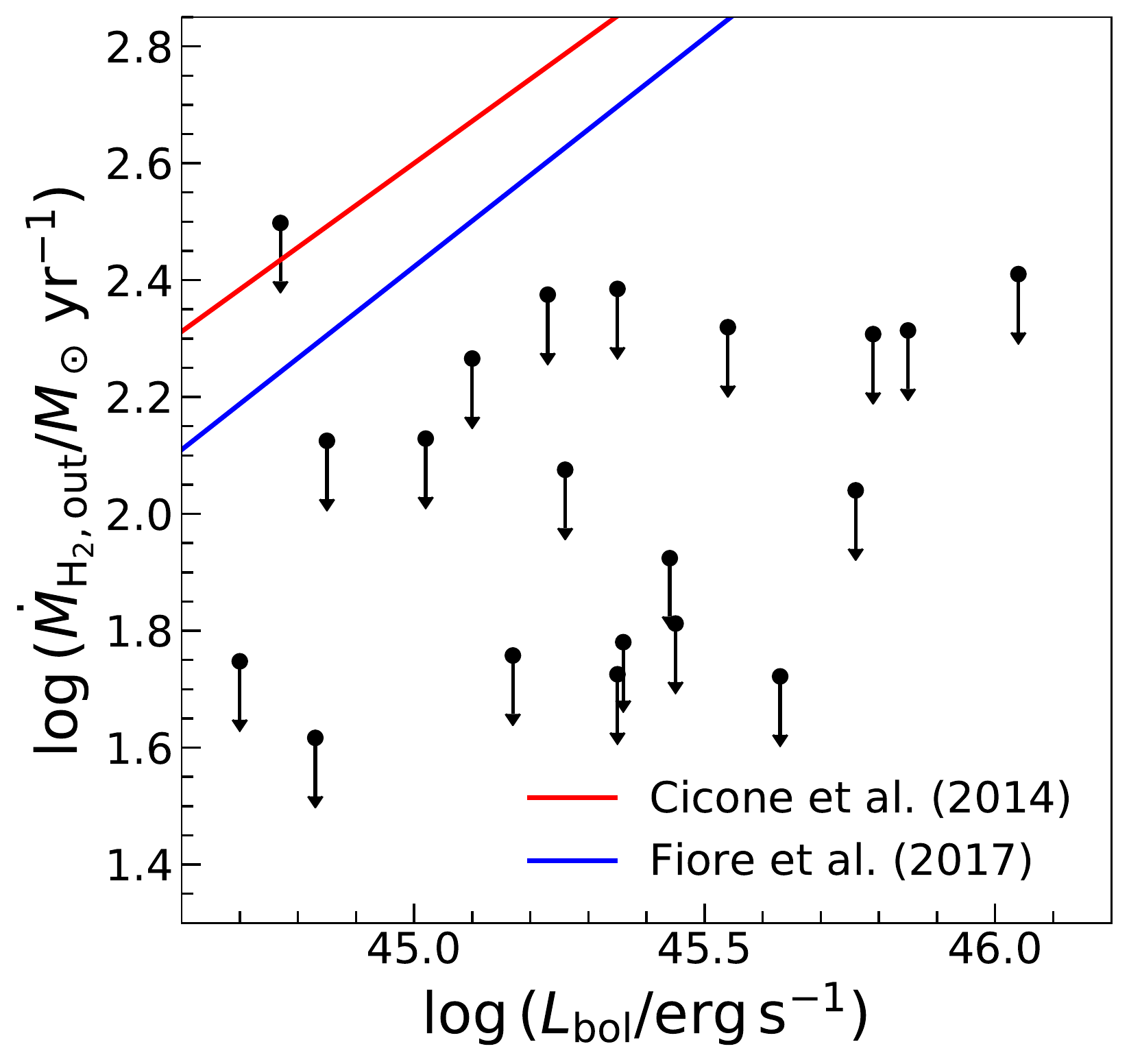}
\caption{Upper limits of the mass outflow rate are plotted against the AGN bolometric luminosity.  PG quasars have much weaker molecular outflows, if any, compared to the relations found by \cite{Cicone2014AA} and \cite{Fiore2017AA}.  The molecular gas masses of the outflow were calculated using $\aco=0.8\,\uaco$, following the same convention as \cite{Cicone2014AA} and \cite{Fiore2017AA}.  Even if we adopt the higher value of $\aco=3.1\,\uaco$, which we have assumed throughout the rest of this work, the upper limits of many quasars still would lie below the blue line.}
\label{fig:of}
\end{center}
\end{figure}

We need the line ratio $R_{21} \equiv L^\prime_{\mbox{\scriptsize CO(2--1)}}/L^\prime_{\mbox{\scriptsize CO(1--0)}}$ and CO-to-H$_2$ conversion factor (\aco) to obtain the molecular gas mass from the CO(2--1) luminosity.  Both quantities are highly uncertain for outflows \citep{Lutz2020AA}.  We adopt $R_{21}=0.62$, derived from the integrated CO(2--1) and CO(1--0) emission of quasar host galaxies \citep{Shangguan2020ApJS}, under the assumption that the CO excitation of the outflow is the same as that of the molecular gas in the disk.  While it is still not clear how common optically thin CO outflows are, $R_{21}$ could be $>1$ in this situation (e.g., \citealt{Dasyra2016AA,Cicone2018ApJ,Lutz2020AA}).  Nevertheless, our assumed $R_{21}$ provides a conservative upper limit of \lco\ for the optically thin case.  The value of \aco\ ranges from 0.8 \uaco\ for ultraluminous IR galaxies to 4.3 \uaco\ for the Milky Way.  For example,  \cite{Cicone2018ApJ} combined the CO and \ci\ observations of NGC 6240 and found $\aco=2.1\pm1.2\,\uaco$ in the outflow.  To match the assumptions of \cite{Cicone2014AA} and \cite{Fiore2017AA},  we momentarily change our assumption of \aco\ from 3.1 to 0.8~\uaco\ to estimate the upper limit of the outflow mass.  As Figure~\ref{fig:of} shows, the limits for the outflow rates of PG quasars deviate systematically below the values expected from previously established relations between mass outflow rate and AGN bolometric luminosity \citep{Cicone2014AA,Fiore2017AA}.  We emphasize that the values of the outflow upper limits are highly uncertain, both because of the poorly known value of \aco\ and the choice of the outflow radius.  Larger $R_\mathrm{out}$ leads to lower upper limits on $\dot{M}_\mathrm{H_2, out}$.   We assume, as do \cite{Cicone2014AA}, $R_\mathrm{out}=R_{\scriptsize \mbox{CO(2--1)}}$; this is a reasonable choice, as it is close to the radius of the observed molecular outflows.  Bearing in mind the above uncertainties, our upper limits indicate that the $\dot{M}_\mathrm{H_2, out}-L_\mathrm{bol}$ relations in the literature are likely biased by the current sample of AGNs with strong outflows.  Our results show that most nearby quasars, while abundant in molecular gas, do {\it not}\ drive strong molecular outflows.

\section{Discussion} 
\label{sec:disc}

\subsection{Gas Fraction and AGN Properties}
\label{ssec:gasagn}

\begin{figure*}
\begin{center}
\includegraphics[height=0.3\textheight]{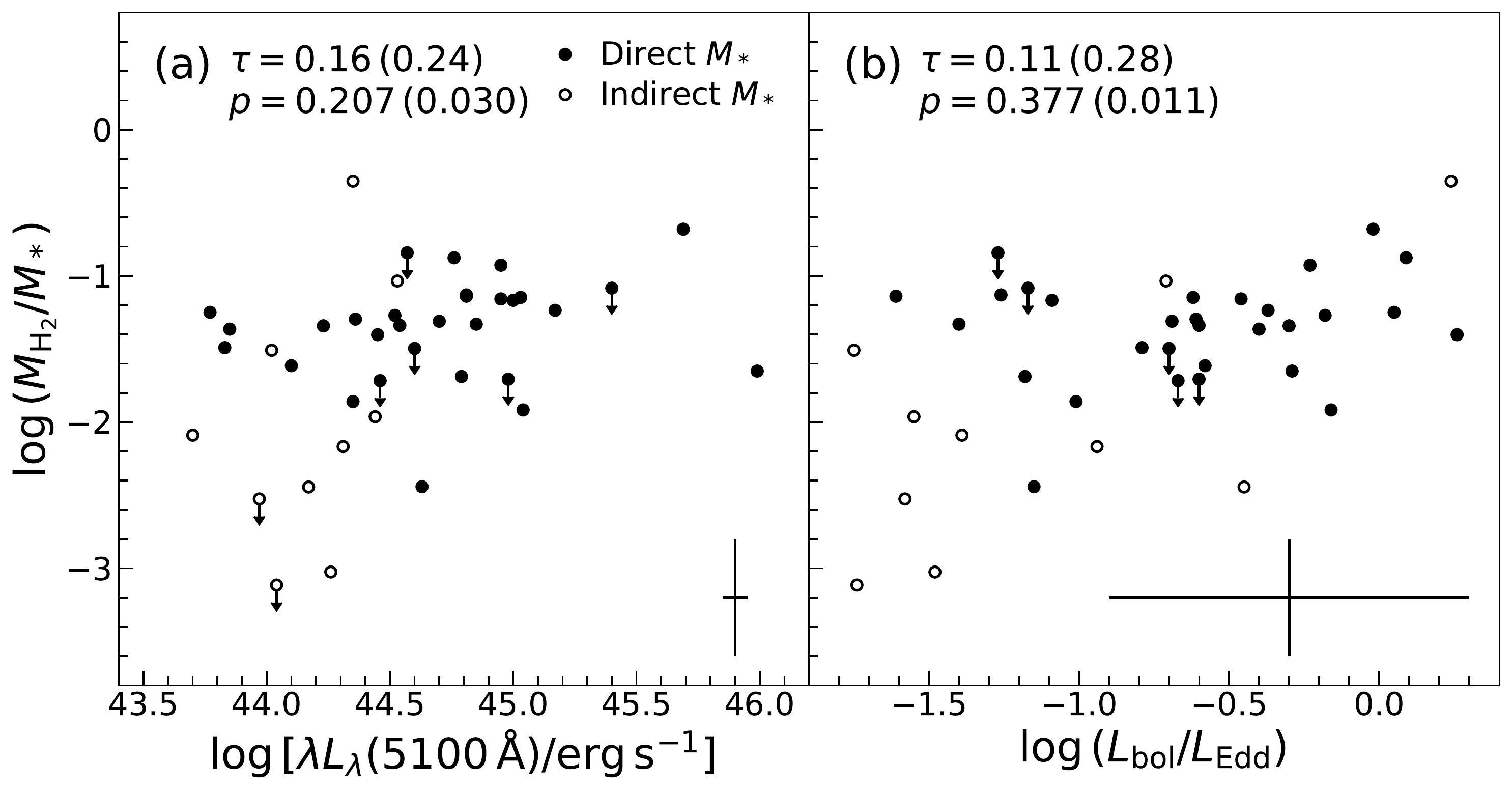}
\caption{The dependence of molecular gas fraction ($\mmol/M_*$) on (a) \lagn\ and (b) Eddington ratio.  Stellar masses derive from direct estimates using high-resolution near-IR images from \citet[][filled circles]{Zhang2016ApJ} and from indirect estimates using the $M_{\rm BH}-M_*$ relation of early-type galaxies from (\citealt{Greene2020}; open circles).  The generalized Kendall's correlation coefficient ($\tau$) and the corresponding $p$-value, accounting for the upper limits, are shown on the upper-left corner of each panel.  The first set of numbers is based on the subsample with direct stellar masses, while the numbers in parentheses are based on the entire sample.  The typical uncertainties of the data are shown on the lower-right corner of each panel.  The y-axis uncertainty corresponds to the subsample with direct stellar masses.}
\label{fig:agn}
\end{center}
\end{figure*}

\cite{Izumi2018PASJ}, analyzing archival CO observations of 37 low-redshift quasars, mostly derived from the PG and Hamburg/ESO \citep{Wisotzki2000AA} surveys, reported a tentative correlation between molecular gas fraction ($\mmol/M_*$) and Eddington ratio.  More than one-third of the Izumi sample only have CO upper limits.  We revisit this problem with our sample, which is similar in size yet more sensitive on account of the new ALMA observations (translating to fewer upper limits).  As shown in Figure \ref{fig:agn}, molecular gas fraction does not correlate significantly with either AGN luminosity or Eddington ratio.  This is particularly true if we only focus on the subsample with direct stellar masses.  Meanwhile, the entire sample shows moderately significant correlations between gas fraction and both AGN luminosity and Eddington ratio.  This is mainly driven by the objects with indirect stellar masses, which tend to have relatively low luminosity and Eddington ratio.  Given the large uncertainty (0.65 dex) of the indirect stellar masses, we regard these moderately significant correlations as suggestive but highly tentative.  

We note that while we performed our correlation analysis,  as did \cite{Izumi2018PASJ}, using the generalised Kendall's $\tau$ test, our implementation of the test with the \texttt{cenken} function yields lower $\tau$ and higher $p$-value than the {\tt IRAF.STSDAS} task \texttt{bhkmethod} used by Izumi.  The latter is likely less robust (E. D. Feigelson 2020, private communications).

\subsection{Star Formation in Quasar Host Galaxies}
\label{ssec:sfr}

Since the AGN does not substantially contribute to the IR luminosity of the host galaxy (Section~\ref{sec:liragn}), we can safely use the IR luminosity to infer the SFR.  From Kennicutt's (1998) calibration, after reducing the original normalization by a factor of 1.5 to convert to a \cite{Kroupa2001MNRAS} stellar initial mass function \citep{Madau2014ARAA},
 
\begin{equation}\label{eq:sfr}
\mathrm{SFR}(\usfr) = 3\times 10^{-44} L_\mathrm{IR}(\ergs).
\end{equation}

\begin{figure}
\begin{center}
\includegraphics[height=0.3\textheight]{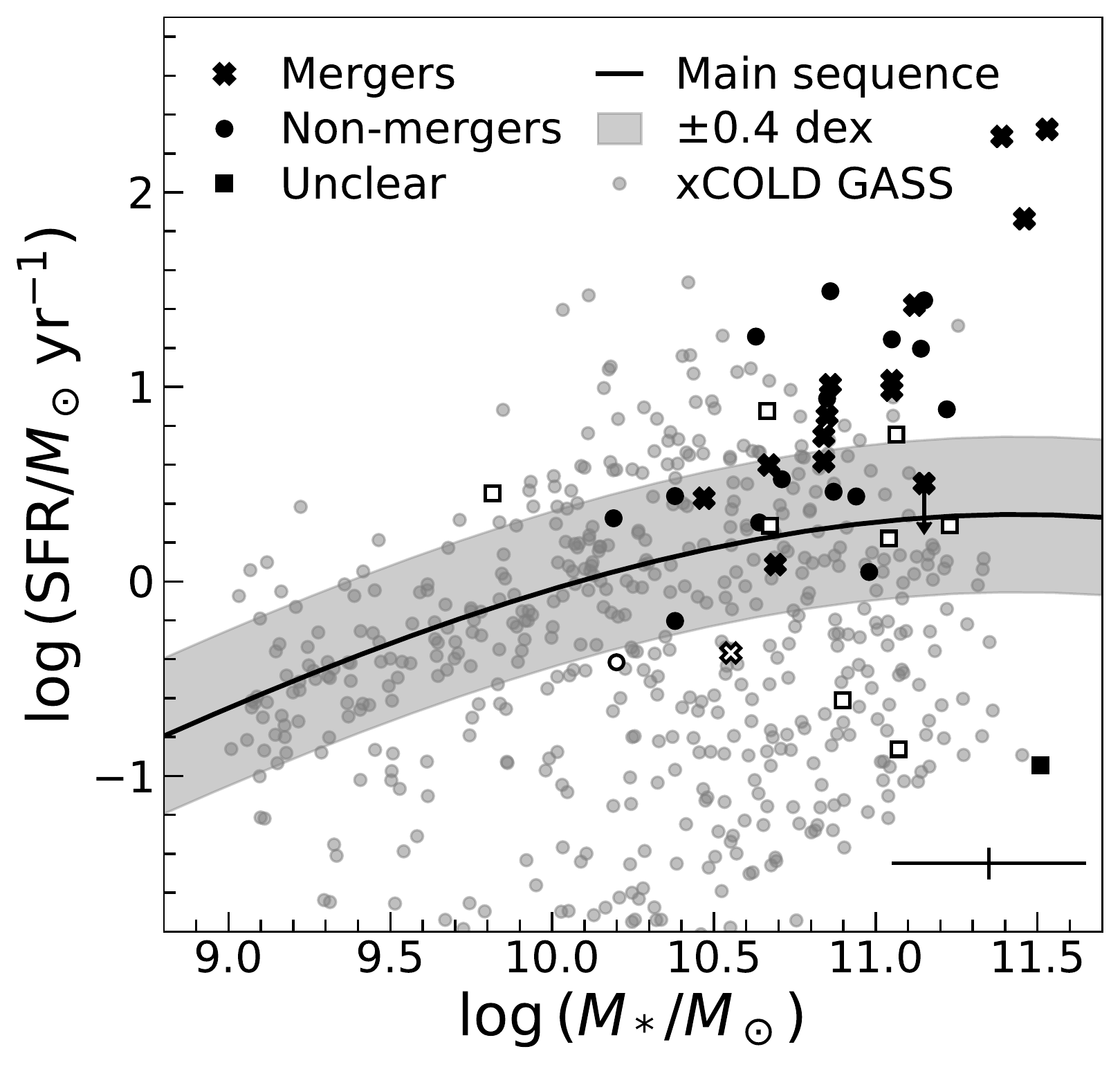}
\caption{The quasar host galaxies are mostly on or above the main sequence of star-forming galaxies, which is denoted by black solid curve and the shaded region ($1\,\sigma$ scatter).  The morphologies of the host galaxies are classified into mergers (crosses), non-mergers (circles), and unclear (squares).  The filled symbols denote the quasars with directly measured stellar masses, while the open symbols denote those with indirect stellar masses.  The grey circles are the inactive galaxies from the xCOLD GASS sample \citep{Saintonge2017ApJS}.  The typical uncertainties are plotted on the lower-right corner.  For clarity, the uncertainty of the vertical axis has been increased by a factor of 3.}
\label{fig:ms}
\end{center}
\end{figure}

\noindent
As shown in Figure~\ref{fig:ms}, quasar host galaxies lie mostly on or above the ``main sequence'' of star-forming galaxies (e.g., \citealt{Peng2010ApJ,Saintonge2017ApJS}).  A main sequence galaxy with stellar mass $M_* \approx 10^{10.5}-10^{11.5}\,M_\odot$, which is characteristic of most of our quasar hosts, has ${\rm SFR} \approx 1\,\usfr$.  By comparison, the SFRs of our quasar hosts range from $\sim 0.1$ to 200 \usfr, with a median value of $\sim 4$ \usfr.  Three sources fall well below the main sequence.  The IR luminosity of PG~1226+023 (3C~273) is highly uncertain because its far-IR spectral energy distribution is dominated by the AGN torus and jet \citep{Shangguan2018ApJ, Zhuang2018}.  The other two (PG~0049+171 and PG~2304+042) are the only sources not detected in our ALMA CO survey.  

Examining the stellar morphologies of the galaxies reveals an unexpected puzzle.  While the majority of the hosts identified as mergers do indeed lie above the main sequence---the three objects in the sample with SFR $\gtrsim\,100\,M_\odot\,{\rm yr}^{-1}$ are all mergers---evidently not all hosts above the main sequence can be classified as such.  These conclusions still hold if we discount the host galaxies with companions as mergers.  Of the 19 sources that formally lie above the $1\,\sigma$ scatter of the main sequence boundary defined by Saintonge et al. (2017), seven (37\%) are classified as non-mergers.

\begin{figure}
\begin{center}
\includegraphics[height=0.3\textheight]{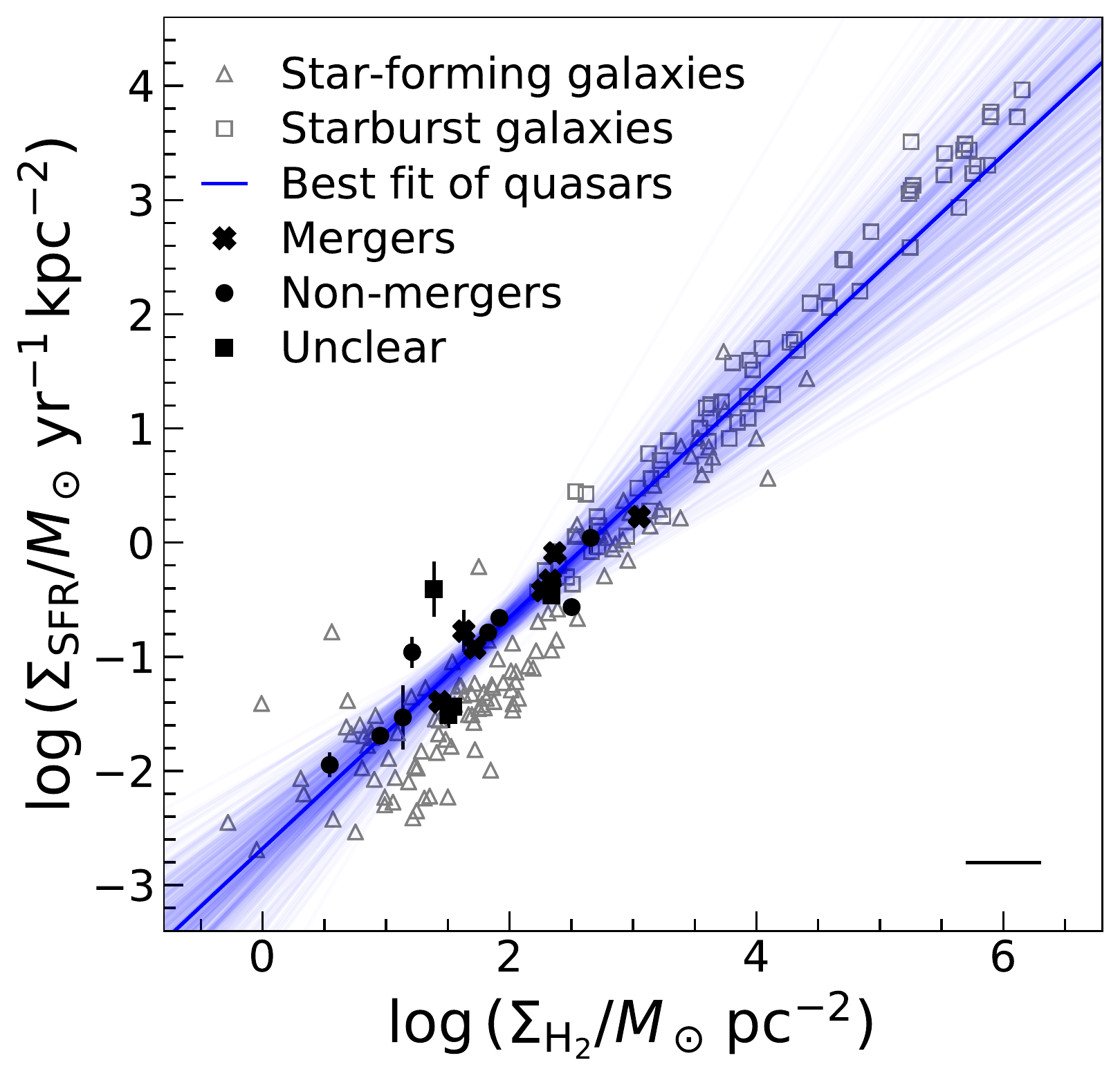}
\caption{The surface density of molecular gas mass and SFR of quasar host galaxies follow a similar trend as star-forming galaxies (grey triangles) and starburst galaxies (grey squares).  The morphologies of the host galaxies are classified into mergers (crosses), non-mergers (circles), and unclear (squares).  The blue solid line is the best-fit relation for quasars, with the faint blue lines indicating the uncertainty of the fit.  The uncertainty of the SFR surface density is shown for each quasar.  The typical uncertainty of the molecular gas surface density, which is representative for all the targets, is plotted on the lower-right corner.
}
\label{fig:kslaw}
\end{center}
\end{figure}

The ALMA subsample affords us the opportunity to calculate the surface density of the SFR and molecular gas mass. We assume, for simplicity, that the physical scale of the star-forming region is equal to the size of the CO emission.  The tight relation between SFR surface density and molecular gas mass surface density obeyed by star-forming and starburst galaxies \citep{Kennicutt1998ApJ,Bigiel2008AJ,Leroy2013AJ} extends to quasar host galaxies (Figure~\ref{fig:kslaw}),
 
\begin{equation}
\log\,\Sigma_\mathrm{SFR} = \left(1.01^{+0.17}_{-0.14}\right) \log\,
\Sigma_\mathrm{H_2} -2.69\left(^{+0.35}_{-0.28}\right),
\end{equation}

\noindent 
where $\Sigma_\mathrm{SFR} \equiv \frac{\mathrm{SFR}}{\pi R_\mathrm{CO}^2}$ and $\Sigma_\mathrm{H_2} \equiv \frac{M_\mathrm{H_2}}{\pi R_\mathrm{CO}^2}$.  The scatter of the data around the best-fit relation ($\sim 0.28$ dex) is dominated by the uncertainties of the measurements.  The points for the comparison sample of star-forming galaxies (grey triangles) and starburst galaxies (grey squares) in Figure~\ref{fig:kslaw} were derived using the \lco, \lir, and diameter measurements published by \cite{Liu2015ApJ}.  For consistency with this study, all SFRs are based on Equation~\ref{eq:sfr}, and the molecular masses assume $\aco=3.1\,\uaco$.

It is clear that quasar host galaxies follow the ``molecular Kennicutt-Schmidt law'' \citep{Schmidt1959ApJ,Kennicutt1998ApJ}.  As in star-forming galaxies (e.g., \citealt{Bigiel2008AJ}), the slope is $\sim 1$.  However, the normalization seems more consistent with that of starburst galaxies instead of normal star-forming galaxies, although the absolute values of $\Sigma_\mathrm{SFR}$ and $\Sigma_\mathrm{H_2}$ are much lower than those of starbursts.  There is much debate as to whether starbursts and normal star-forming galaxies share the same value of \aco\ (e.g., \citealt{Genzel2010MNRAS,Bolatto2013ARAA,Liu2015ApJ}), but a discussion of this topic is beyond the scope of this paper.

\begin{figure*}
\begin{center}
\includegraphics[height=0.3\textheight]{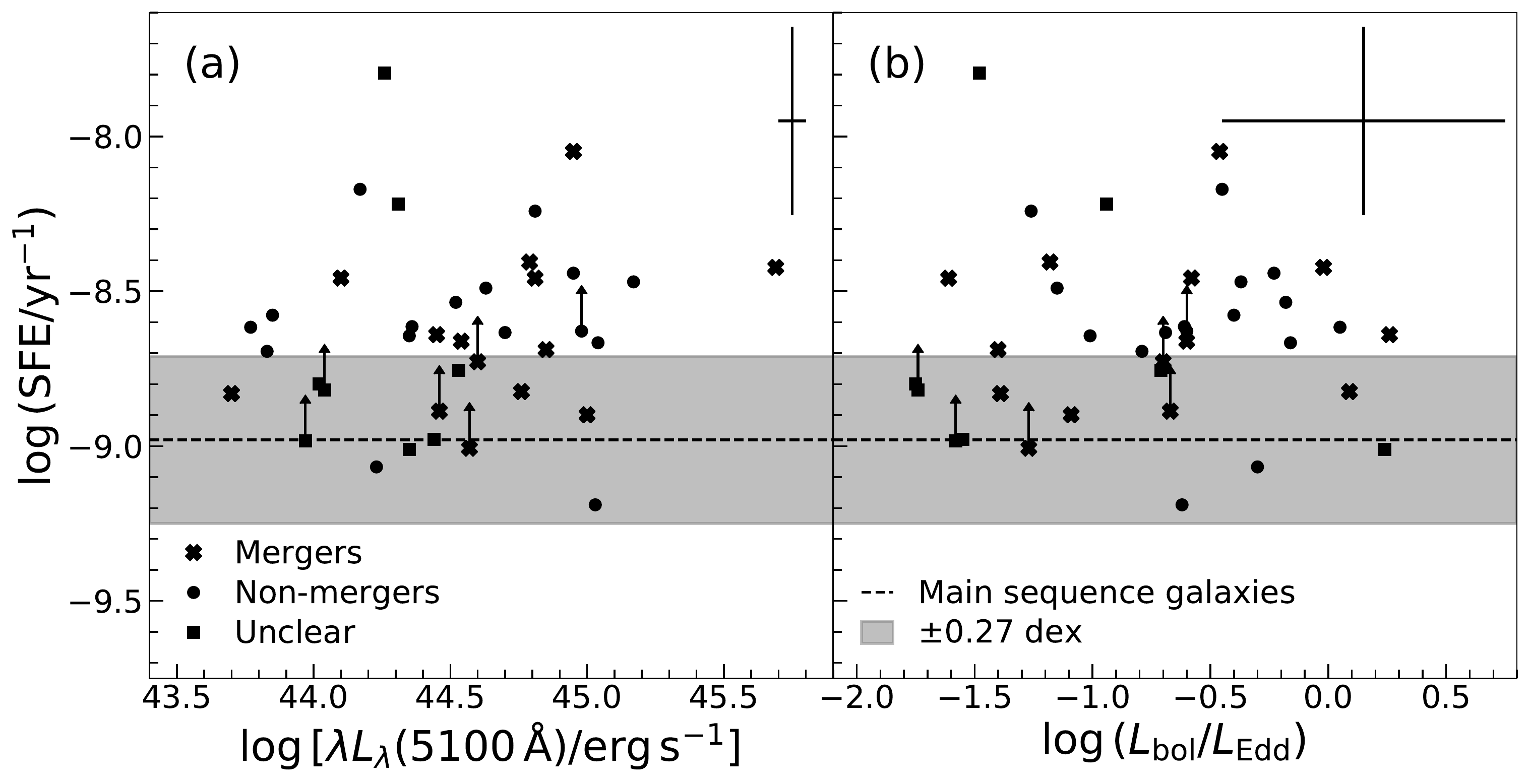}
\caption{The star formation efficiency (SFE) of the quasar host galaxies are plotted against (a) the AGN 5100~\AA\ continuum luminosity and (b) Eddington ratio.  The morphologies of the host galaxies are classified into mergers (crosses), non-mergers (circles), and unclear (squares).  Typical uncertainties are plotted on the upper-right corner of each panel.  The dashed horizontal line and shaded region indicate the average SFE and its scatter of the star-forming galaxies on the main sequence \citep{Saintonge2017ApJS}.  Quasar host galaxies show systematically higher SFE than inactive galaxies.  However, the SFE is not correlated with either the AGN luminosity or Eddington ratio, for the sample as a whole or for individual subsamples.  
}
\label{fig:sfe}
\end{center}
\end{figure*}

The SFE does not depend on quasar luminosity or Eddington ratio, either for the entire sample or for subsamples of different morphologies (Figure~\ref{fig:sfe}).  Taken at face value, the above results imply that quasar host galaxies form stars more efficiently than main sequence star-forming galaxies.  This is simply another expression of the \lir--\lco\ relation, already reported in \cite{Shangguan2020ApJS}.  The exact normalization of the $\Sigma_\mathrm{SFR} - \Sigma_\mathrm{H_2}$ relation for our sample may be underestimated if the star-forming regions of quasar hosts have complex structures that are much smaller than the size of the globally measured CO emission.  While detailed observation and analysis are needed, complex structures are revealed with high-resolution ALMA observations of several quasars in our sample (J. Molina et al. in preparation).

It is still an open question as to why the quasar host galaxies are starbursts.  Positive AGN feedback has been invoked to account for star formation activity in both $z \approx 2$ quasar host galaxies (\citealt{Cresci2015ApJ,Carniani2016AA}; but see \citealt{Scholtz2020MNRAS} for counterarguments) and nearby AGNs \citep{Maiolino2017Natur,Gallagher2019MNRAS}.  Our partial correlation analysis of PG quasars (Figure \ref{fig:ir51}; Section~\ref{sec:liragn}), however, suggests that the AGN does not further enhance the SFR significantly, after the common dependence between AGN luminosity and SFR on the molecular gas is removed.  This result needs to be confirmed with a much larger sample.  In the mean time, we cannot rule out the possibility that positive AGN feedback enhances the SFR at a modest ($\sim 10\%$) level, given the 0.3 dex intrinsic scatter for the \lir--\lco\ relation.  It would be instructive to apply the same partial correlation test to high-redshift quasars to see whether outflow-driven star formation plays a more dominant role in these more powerful systems.

\section{Summary}
\label{sec:sum}

We combine our new ALMA CO(2--1) survey \citep{Shangguan2020ApJS} with measurements from the literature to investigate the molecular gas properties of 40 low-redshift quasars that form a representative subset of the parent sample of PG quasars at $z<0.3$.  This is the largest and most sensitive study of molecular gas emission to date for nearby quasars.  We compare the molecular gas masses and kinematics of our sample with those of local inactive galaxies to evaluate the nature of star formation and AGN feedback in quasar host galaxies.

We report the following findings:

\begin{itemize}
\item The molecular gas masses of most low-redshift quasar host galaxies are consistent with those of galaxies on the star-forming main sequence.  Only 20\% of the quasar hosts are gas-poor ($\mmol/M_* \lesssim 0.01$).
 
\item The CO line exhibits kinematically regular profiles, whose deprojected line widths yield rotation velocities consistent with the CO Tully--Fisher relation of star-forming galaxies.

\item Despite the coexistence of abundant molecular gas and powerful quasar activity, no obvious high-velocity CO emission from molecular gas outflows is detected.  We calculate conservative upper limits of the mass outflow rate, which lie systematically and markedly below an empirical relation between mass outflow rate and AGN luminosity previously established from AGNs with detected molecular outflows.  
\item Consistent with previous works, CO luminosity correlates significantly with AGN luminosity but not Eddington ratio.  AGN luminosity is correlated with \lco\ and $M_{\rm BH}$, strongly suggesting that AGN fueling is coupled to the cold gas reservoir of the host galaxy.
 
\item The molecular gas mass fraction ($\mmol/M_*$) does not significantly depend on \lagn\ or Eddington ratio.

\item We show that the observed strong relation between the global IR luminosity (\lir) and AGN luminosity [\lagn] is driven mainly by their mutual dependence on \lco.  No significant partial correlation exists between \lir\ and \lagn\ after removing their dependence on \lco.  This implies that \lir\ for this sample of low-redshift quasars does not suffer from appreciable contamination from AGN heating, and hence can be used to estimate the SFR for the host galaxy.

\item Quasar host galaxies have an enhanced SFE similar to starburst galaxies, as evidenced by their location on the Kennicutt-Schmidt relation and position above the main sequence of star-forming galaxies, but the SFE shows no correlation with AGN luminosity or Eddington ratio.  
 
\item Mergers do not appear to be a necessary condition for enhancing the SFE in quasar hosts.

\end{itemize}

The above findings paint a highly nuanced picture of BH--galaxy coevolution.  On the one hand, we find that the cold gas supply is the common ingredient that ties together BH accretion and star formation in the host galaxy.  On the other hand, although our study specifically targets unobscured AGNs powerful enough to be considered quasars, we find only scant evidence that ``quasar-mode'' feedback exerts any impact on the content or kinematics of the cold gas.  As in our earlier study using gas masses inferred indirectly from dust masses \citep{Shangguan2018ApJ}, the CO measurements reported here directly confirm that the host galaxies of nearby quasars generally are far from gas-poor.  Not only do they have abundant molecular gas, but the gas resides in a kinematically regular disk, as evidenced by their adherence to the CO Tully--Fisher relation of inactive galaxies.  The integrated profiles look normal, too, showing no sign of high-velocity wings.   Far from quenched, the star formation activity of nearby quasars actually surpasses that of main sequence galaxies of comparable stellar mass and gas supply.  A significant fraction of the quasar hosts can be regarded as starburst galaxies, but merger signatures are not universally present.

\acknowledgments

We thank the anonymous referee for helpful suggestions.  We acknowledge support from the National Science Foundation of China grant 11721303 and 11991052 (LCH), the National Key R\&D Program of China grant 2016YFA0400702 (LCH), CONICYT-Chile grants Basal AFB-170002 (FEB, ET), FONDO ALMA 31160033 (FEB), FONDECYT Regular 1160999 (ET), 1200495 (FEB, ET) and 1190818 (ET, FEB), and Anillo de ciencia y tecnologia ACT1720033 (ET), and the Chilean Ministry of Economy, Development, and Tourism's Millennium Science Initiative through grant IC120009, awarded to The Millennium Institute of Astrophysics, MAS (FEB).  JS thanks Eric Feigelson and Hassen Yusef for valuable advice on statistical methods, Yulin Zhao for sharing the \galfit\ results of the PG quasar host galaxies, Juan Molina for sharing the high-resolution CO measurements, and Yanxia Xie, Ming-Yang Zhuang, and Hagai Netzer for helpful discussions.

\facilities{ALMA}

\software{astropy \citep{Astropy2013AA}, ASURV \citep{Feigelson1985ApJ}, CASA \citep{McMullin2007ASP}, NADA \citep{Lee2017}}

\appendix
\section{Testing the CO size with simulated data}
\label{apd:size}

\begin{figure}
\begin{center}
\includegraphics[height=0.3\textheight]{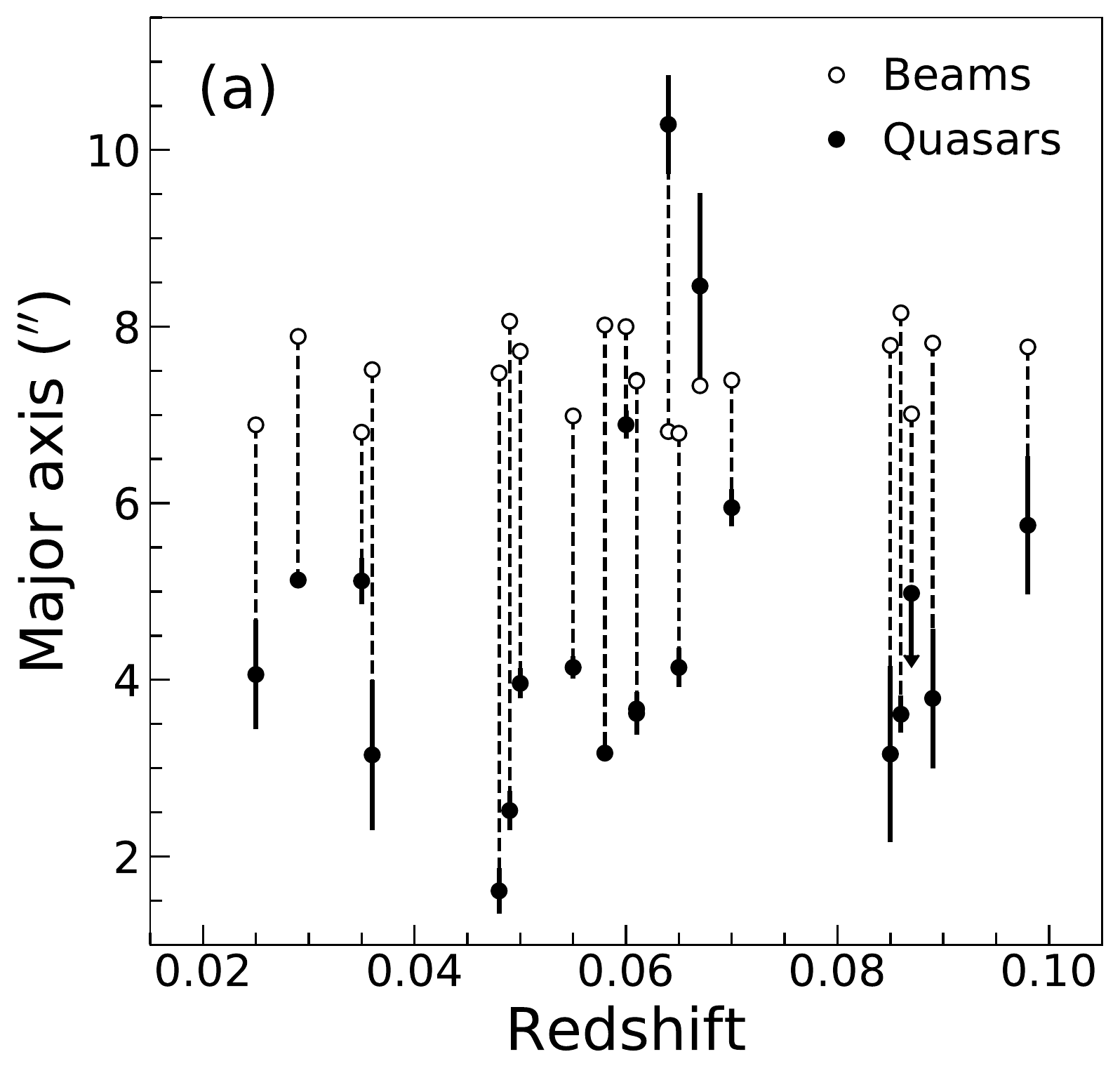}
\includegraphics[height=0.3\textheight]{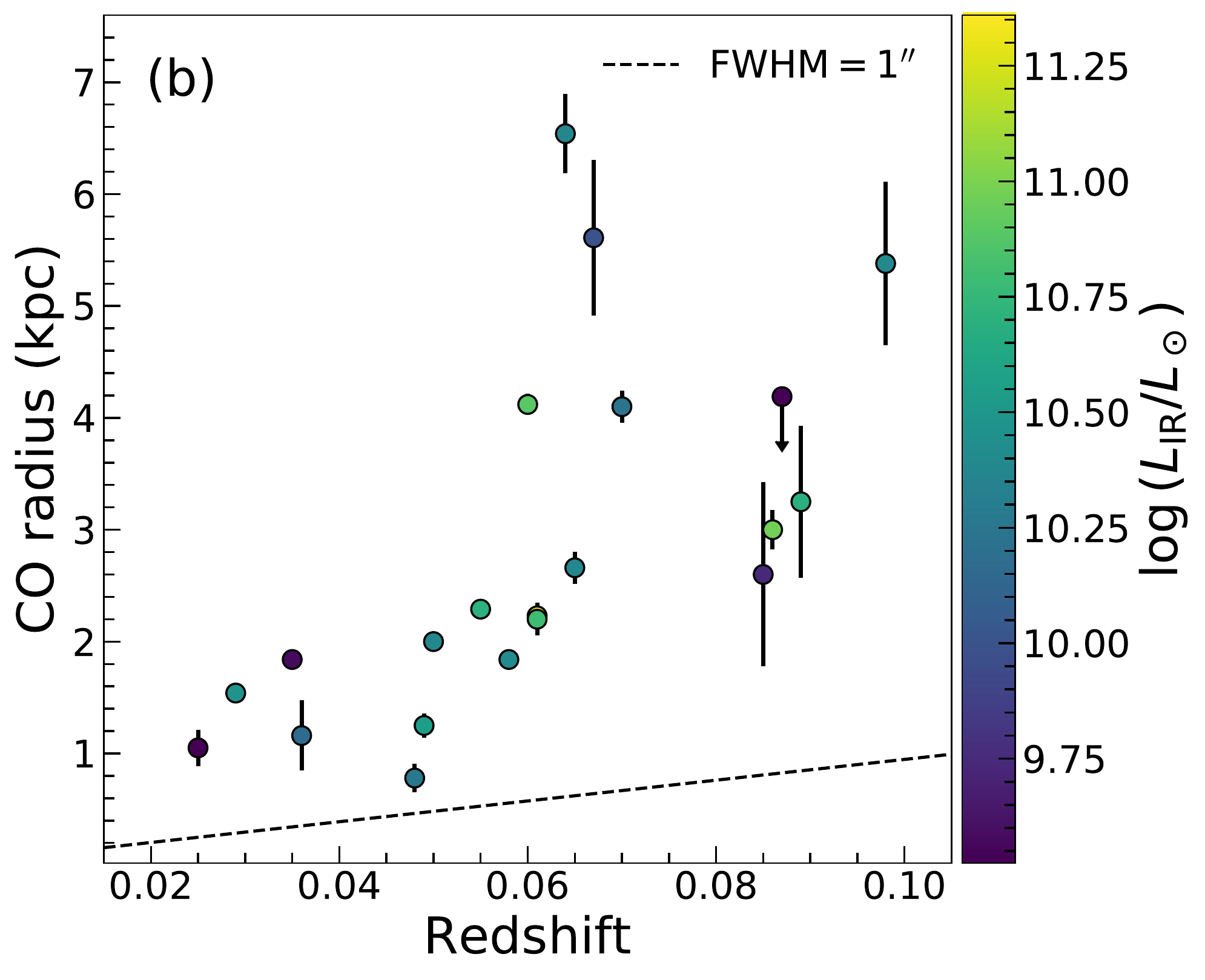}
\caption{(a) Comparison of the major axes of the beams (open circles) and the CO emission of the quasars (filled circles) that are measured by fitting the $uv$ data.  The vertical dashed line connects the same quasar.  Most of the CO sizes are smaller than the corresponding beam sizes.  (b) The physical radii of the CO emission of the quasars are above the resolution limit ($\mathrm{FWHM}=1\arcsec$), below which \texttt{uvmodelfit} cannot derive a robust size.  The IR luminosity of the host galaxy does not show a clear correlation with the CO size.}
\label{fig:size}
\end{center}
\end{figure}

We fit the $uv$ data using  \texttt{uvmodelfit} to derive the size of the CO emission, which we take as the FWHM of the two-dimensional Gaussian model.  The CO sizes, which are well-constrained by the high signal-to-noise ratio of the $uv$ data, are usually smaller than the beam size of the corresponding observations (Figure~\ref{fig:size}a).  The quasars at $z \gtrsim 0.06$ show larger CO sizes than those at lower redshift (Figure~\ref{fig:size}b).  This trend is not correlated with the IR or CO luminosity.  Despite the relatively large synthesis beam, the size measurements are robust because they are larger than the resolution limit.

To test whether fitting the $uv$ data can yield reliable sizes, and to ascertain the limit to which sizes can be extracted from our observations, we simulated our observations using the CASA task \texttt{simobserve}, using configuration parameters appropriate for our Cycle 5 ACA observations.  The input models are two-dimensional Gaussian profiles with a total flux density of 1~Jy at 230~GHz and FWHM 0.5\arcsec--6.0\arcsec, axis ratio $\sim 0.7$, and position angle 47\degree.  The integration time is set to $\sim 2.5$~hours.  These are typical values derived from the real data (Table~\ref{tab:of}).  As we are concerned only with the size estimates derived from the $uv$ data, we assume the same Gaussian profile across the 0.5~GHz bandwidth.  
Thermal noise (``tsys-atm'') is assumed in the simulation, although the results are not sensitive to whether thermal noise is included.  Including a more realistic noise level is challenging. Fortunately, the high signal-to-noise of our observations renders the treatment of noise secondary.

\begin{figure}
\begin{center}
\includegraphics[height=0.3\textheight]{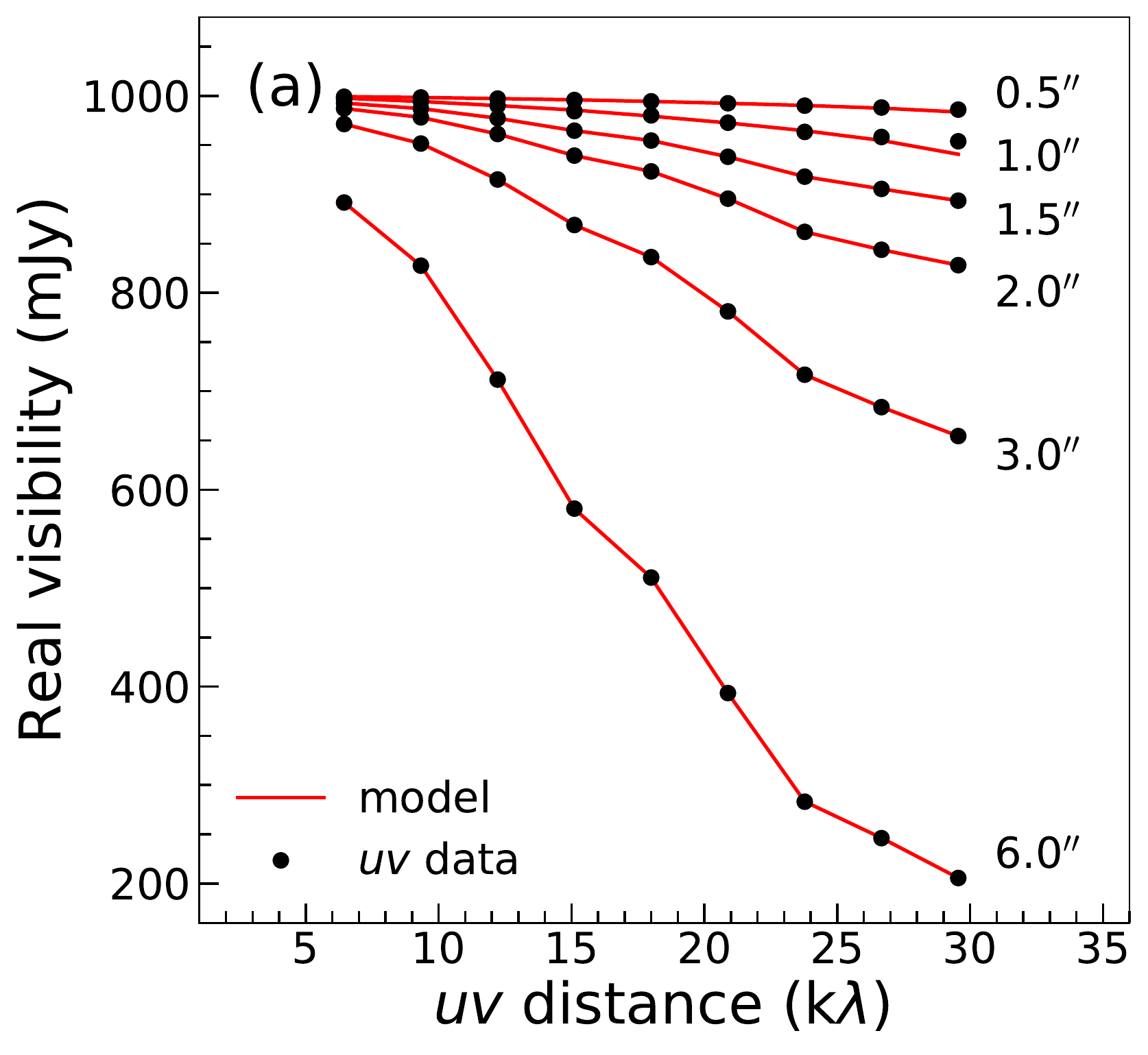}
\includegraphics[height=0.3\textheight]{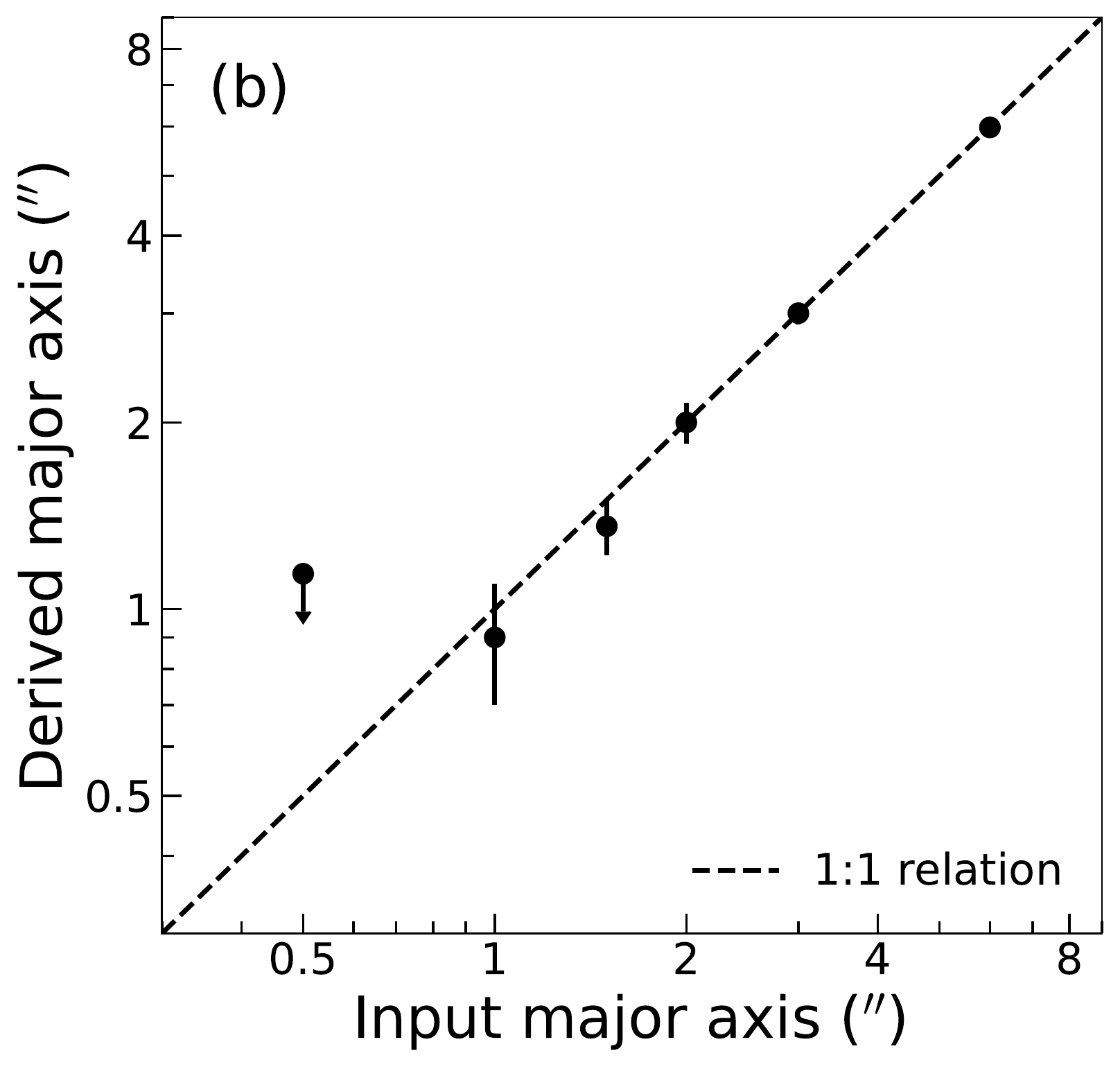}
\caption{(a) The real part of the complex visibility of the simulated data (filled circles).  The FWHM of the major axis of the Gaussian profiles are labeled on the right.  The red curves are the best-fit models from \texttt{uvmodelfit}.  (b) The FWHM of the major axis derived from \texttt{uvmodelfit} is consistent with the input model for sizes larger than 1\arcsec.}
\label{fig:sim}
\end{center}
\end{figure}

We fit the $uv$ data of the simulated observations with \texttt{uvmodelfit} (Figure~\ref{fig:sim}a), using the same Gaussian model as described in Section~\ref{ssec:size}.  Since the position angle becomes quite uncertain for sizes $\lesssim 1\farcs5$, we fix the axis ratio to unity, and, in view of the large uncertainty of sizes $\lesssim 1$\arcsec, we consider the size to be an upper limit when the input size is 0\farcs5.
As illustrated in Figure~\ref{fig:sim}b, we conclude that \texttt{uvmodelfit} yields robust size measurements for our Cycle 5 ACA data when the emission size is $\gtrsim 1\arcsec$.

\input{tab1.tex}

\input{tab2.tex}

\end{document}

%% file: tab1.tex

\begin{deluxetable*}{c c c l c c c c c c l c c}
\tablecaption{Molecular Gas Properties of PG Quasars\label{tab:phy}}
\tabletypesize{\tiny}
\tablehead{
\colhead{Object} &
\colhead{$\log\,\lambda L_\lambda(5100\,\mathrm{\AA})$} &
\colhead{$\log\,M_\mathrm{BH}$} &
\colhead{$\log\,M_*$} &
\colhead{$\log\,L_\mathrm{IR}$} &
\colhead{$q$} &
\colhead{Ref.} &
\colhead{$i$} &
\colhead{$\log\,L^\prime_{\tiny\mbox{CO(1--0)}}$} &
\colhead{$\log\,M_\mathrm{H_2}$} &
\colhead{$W_{50}$} &
\colhead{Morphology} &
\colhead{Ref.} \\
\colhead{} &
\colhead{($\mathrm{erg\,s^{-1}}$)} &
\colhead{($M_\odot$)} &
\colhead{($M_\odot$)} &
\colhead{($\mathrm{erg\,s^{-1}}$)} &
\colhead{} &
\colhead{} &
\colhead{($\degree$)} &
\colhead{($\mathrm{K\,km\,s^{-1}\,pc^2}$)} &
\colhead{($M_\odot$)} &
\colhead{($\mathrm{km\,s^{-1}}$)} &
\colhead{}
}
\colnumbers
\startdata
 \mcl{12}{c}{ALMA Sample} \\ \hline
 PG 0003$+$199 & 44.17 & 7.52 &      10.20\tnma & $43.11^{+0.03}_{-0.03}$ &    0.93 &      1  &   22.03 &       7.26$\pm$0.07 &        7.75$\pm$0.31 & $155.06^{+16.37}_{-14.67}$ &       D &          1  \\
 PG 0007$+$106 & 44.79 & 8.87 &           10.84 & $44.27^{+0.02}_{-0.03}$ & \nodata & \nodata & \nodata &       8.66$\pm$0.03 &        9.15$\pm$0.30 & $386.78^{+29.57}_{-25.18}$ &       M &          3  \\
 PG 0049$+$171 & 43.97 & 8.45 &      10.90\tnma & $42.91^{+0.05}_{-0.08}$ & \nodata & \nodata & \nodata & \mcl{1}{c}{$<7.88$} &  \mcl{1}{c}{$<8.37$} &        \mcl{1}{c}{\nodata} & \nodata &     \nodata \\
 PG 0050$+$124 & 44.76 & 7.57 &           11.12 & $44.94^{+0.01}_{-0.01}$ &    0.53 &      2  &   60.27 &       9.75$\pm$0.01 &       10.24$\pm$0.30 &   $377.77^{+0.85}_{-0.86}$ &     D,c &          2  \\
 PG 0923$+$129 & 43.83 & 7.52 &           10.71 & $44.05^{+0.01}_{-0.02}$ &    0.78 &      2  &   39.37 &       8.73$\pm$0.01 &        9.22$\pm$0.30 &   $361.68^{+1.07}_{-1.03}$ &       D &          2  \\
 PG 0934$+$013 & 43.85 & 7.15 &           10.38 & $43.96^{+0.02}_{-0.02}$ &    0.69 &      2  &   48.03 &       8.52$\pm$0.03 &        9.02$\pm$0.30 &   $217.84^{+7.98}_{-7.15}$ &       D &          2  \\
 PG 1011$-$040 & 44.23 & 7.43 &           10.87 & $43.98^{+0.02}_{-0.02}$ &    0.92 &      2  &   24.27 &       9.04$\pm$0.01 &        9.53$\pm$0.30 &   $141.00^{+1.42}_{-1.35}$ &       D &          2  \\
 PG 1119$+$120 & 44.10 & 7.58 &           10.67 & $44.12^{+0.02}_{-0.04}$ &    0.63 &      2  &   52.31 &       8.56$\pm$0.02 &        9.06$\pm$0.30 &   $212.68^{+2.41}_{-2.37}$ &     D,c &          2  \\
 PG 1126$-$041 & 44.36 & 7.87 &           10.85 & $44.46^{+0.03}_{-0.03}$ & \nodata & \nodata & \nodata &       9.06$\pm$0.01 &        9.55$\pm$0.30 &   $467.00^{+1.74}_{-1.72}$ &       D &          4  \\
 PG 1211$+$143 & 45.04 & 8.10 &           10.38 & $43.32^{+0.05}_{-0.05}$ &    0.84 &      1  &   33.63 &       7.97$\pm$0.04 &        8.46$\pm$0.30 &    $65.90^{+7.29}_{-7.00}$ &       D &          1  \\
 PG 1229$+$204 & 44.35 & 8.26 &           10.94 & $43.96^{+0.01}_{-0.01}$ &    0.55 &      1  &   58.47 &       8.59$\pm$0.03 &        9.08$\pm$0.30 &   $202.21^{+3.14}_{-2.84}$ &       D &          1  \\
 PG 1244$+$026 & 43.77 & 6.62 &           10.19 & $43.85^{+0.02}_{-0.01}$ &    0.70 &      2  &   46.63 &       8.45$\pm$0.02 &        8.94$\pm$0.30 &   $108.94^{+2.94}_{-2.91}$ &       D &          2  \\
 PG 1310$-$108 & 43.70 & 7.99 &      10.55\tnma & $43.16^{+0.02}_{-0.01}$ & \nodata & \nodata & \nodata &       7.97$\pm$0.03 &        8.46$\pm$0.30 &   $204.08^{+7.04}_{-6.33}$ &     D,t &          5  \\
 PG 1341$+$258 & 44.31 & 8.15 &      10.67\tnma & $43.81^{+0.04}_{-0.05}$ & \nodata & \nodata & \nodata &       8.01$\pm$0.10 &        8.51$\pm$0.31 &        \mcl{1}{c}{\nodata} & \nodata &     \nodata \\
 PG 1351$+$236 & 44.02 & 8.67 &      11.06\tnma & $44.28^{+0.01}_{-0.01}$ & \nodata & \nodata & \nodata &       9.06$\pm$0.01 &        9.55$\pm$0.30 &   $340.88^{+1.84}_{-1.81}$ & \nodata &     \nodata \\
 PG 1404$+$226 & 44.35 & 7.01 & \ph{0}9.82\tnma & $43.97^{+0.02}_{-0.02}$ & \nodata & \nodata & \nodata &       8.97$\pm$0.03 &        9.46$\pm$0.30 &   $284.69^{+9.27}_{-8.31}$ & \nodata &     \nodata \\
 PG 1426$+$015 & 44.85 & 9.15 &           11.05 & $44.55^{+0.02}_{-0.02}$ & \nodata &      1  & \nodata &       9.23$\pm$0.02 &        9.72$\pm$0.30 & $343.71^{+10.26}_{-10.16}$ &     D,c &          1  \\
 PG 1448$+$273 & 44.45 & 7.09 &           10.47 & $43.95^{+0.02}_{-0.02}$ &    0.63 &      2  &   52.50 &       8.58$\pm$0.02 &        9.07$\pm$0.30 &   $170.78^{+4.07}_{-4.27}$ &       M &          2  \\
 PG 1501$+$106 & 44.26 & 8.64 &      11.04\tnma & $43.74^{+0.07}_{-0.05}$ & \nodata & \nodata & \nodata &       7.52$\pm$0.05 &        8.02$\pm$0.30 &  $192.68^{+10.41}_{-9.49}$ & \nodata &     \nodata \\
 PG 2130$+$099 & 44.54 & 8.04 &           10.85 & $44.37^{+0.02}_{-0.03}$ &    0.44 &      1  &   66.42 &       9.02$\pm$0.01 &        9.51$\pm$0.30 &   $548.36^{+4.98}_{-4.68}$ &     D,t &          1  \\
 PG 2209$+$184 & 44.44 & 8.89 &      11.23\tnma & $43.81^{+0.02}_{-0.03}$ & \nodata & \nodata & \nodata &       8.77$\pm$0.02 &        9.27$\pm$0.30 &   $277.48^{+1.36}_{-1.22}$ & \nodata &     \nodata \\
 PG 2214$+$139 & 44.63 & 8.68 &           10.98 & $43.57^{+0.01}_{-0.02}$ &    0.97 &      2  &   15.14 &       8.05$\pm$0.06 &        8.54$\pm$0.31 &   $179.64^{+9.00}_{-8.89}$ &       E &          2  \\
 PG 2304$+$042 & 44.04 & 8.68 &      11.07\tnma & $42.66^{+0.06}_{-0.08}$ & \nodata & \nodata & \nodata & \mcl{1}{c}{$<7.46$} &  \mcl{1}{c}{$<7.96$} &        \mcl{1}{c}{\nodata} & \nodata &     \nodata \\ \hline
\mcl{12}{c}{Literature Sample} \\ \hline
 PG 0052$+$251 & 45.00 & 8.99 &           11.05 & $44.51^{+0.02}_{-0.02}$ &    0.55 &      2  &   58.35 &    \mcl{1}{c}{9.39} &     \mcl{1}{c}{9.88} &            \mcl{1}{c}{429} &     D,t &          1  \\
 PG 0157$+$001 & 44.95 & 8.31 &           11.53 & $45.85^{+0.03}_{-0.05}$ &    0.60 &      1  &   54.74 &       9.88$\pm$0.04 &       10.37$\pm$0.30 &            \mcl{1}{c}{270} &       M &          6  \\
 PG 0804$+$761 & 45.03 & 8.55 &           10.64 & $43.83^{+0.07}_{-0.05}$ &    0.65 &      2  &   50.50 &       9.00$\pm$0.11 &        9.49$\pm$0.32 &            \mcl{1}{c}{755} &       E &          2  \\
 PG 0838$+$770 & 44.70 & 8.29 &           11.14 & $44.72^{+0.03}_{-0.04}$ & \nodata & \nodata & \nodata &       9.34$\pm$0.07 &        9.83$\pm$0.31 &             \mcl{1}{c}{60} &       D &          4  \\
 PG 0844$+$349 & 44.46 & 8.03 &           10.69 & $43.61^{+0.02}_{-0.02}$ &    0.39 &      1  &   70.02 & \mcl{1}{c}{$<8.48$} &  \mcl{1}{c}{$<8.97$} &        \mcl{1}{c}{\nodata} &       M &          1  \\
 PG 1202$+$281 & 44.57 & 8.74 &           10.86 & $44.53^{+0.03}_{-0.03}$ &    0.92 &      2  &   22.97 & \mcl{1}{c}{$<9.53$} & \mcl{1}{c}{$<10.02$} &        \mcl{1}{c}{\nodata} &     E,c &          2  \\
 PG 1226$+$023 & 45.99 & 9.18 &           11.51 & $42.58^{+0.47}_{-0.57}$ &    0.65 &      2  &   51.04 &       9.37$\pm$0.01 &        9.86$\pm$0.30 &            \mcl{1}{c}{490} &       U &          2  \\
 PG 1309$+$355 & 44.98 & 8.48 &           11.22 & $44.41^{+0.04}_{-0.04}$ & \nodata &      1  & \nodata & \mcl{1}{c}{$<9.02$} &  \mcl{1}{c}{$<9.51$} &        \mcl{1}{c}{\nodata} &       E &          1  \\
 PG 1351$+$640 & 44.81 & 8.97 &           10.63 & $44.78^{+0.04}_{-0.05}$ &    0.98 &      2  &   12.04 &       9.01$\pm$0.08 &        9.50$\pm$0.31 &            \mcl{1}{c}{260} &       E &          2  \\
 PG 1402$+$261 & 44.95 & 8.08 &           10.86 & $45.01^{+0.04}_{-0.04}$ &    0.45 &      1  &   65.71 &    \mcl{1}{c}{9.44} &     \mcl{1}{c}{9.93} &        \mcl{1}{c}{\nodata} &       D &          1  \\
 PG 1411$+$442 & 44.60 & 8.20 &           10.84 & $44.14^{+0.03}_{-0.03}$ &    0.71 &      1  &   45.95 &    \mcl{1}{c}{8.85} &     \mcl{1}{c}{9.34} &        \mcl{1}{c}{\nodata} &       M &          1  \\
 PG 1415$+$451 & 44.53 & 8.14 &      10.67\tnma & $44.40^{+0.02}_{-0.01}$ & \nodata & \nodata & \nodata &       9.14$\pm$0.06 &        9.63$\pm$0.31 &             \mcl{1}{c}{90} & \nodata &     \nodata \\
 PG 1440$+$356 & 44.52 & 7.60 &           11.05 & $44.77^{+0.02}_{-0.01}$ &    0.66 &      1  &   50.06 &       9.29$\pm$0.04 &        9.78$\pm$0.30 &            \mcl{1}{c}{310} &       D &          1  \\
 PG 1444$+$407 & 45.17 & 8.44 &           11.15 & $44.97^{+0.05}_{-0.05}$ &    0.78 &      1  &   39.69 &    \mcl{1}{c}{9.42} &     \mcl{1}{c}{9.91} &            \mcl{1}{c}{257} &       D &          2  \\
 PG 1545$+$210 & 45.40 & 9.47 &           11.15 &                $<44.03$ & \nodata &      1  & \nodata & \mcl{1}{c}{$<9.57$} & \mcl{1}{c}{$<10.07$} &        \mcl{1}{c}{\nodata} &     U,c &          1  \\
 PG 1613$+$658 & 44.81 & 9.32 &           11.46 & $45.39^{+0.02}_{-0.02}$ & \nodata &      1  & \nodata &       9.83$\pm$0.03 &       10.32$\pm$0.30 &            \mcl{1}{c}{400} &       M &          2  \\
 PG 1700$+$518 & 45.69 & 8.61 &           11.39 & $45.81^{+0.02}_{-0.05}$ &    0.49 &      1  &   62.84 &      10.22$\pm$0.08 &       10.71$\pm$0.31 &            \mcl{1}{c}{260} &       M &          2  \\
\enddata
\tablenotetext{a}{The stellar mass is estimated indirectly from the BH mass according to Equation (\ref{eq:ms}) \citep{Greene2020}.}
\tablecomments{
Col.  (1) Source name.
Col.  (2) AGN monochromatic luminosity of the continuum at 5100 \AA.
Col.  (3) BH mass.
Col.  (4) Stellar mass of the host galaxy.  The uncertainties of the direct and indirect stellar mass are $\sim 0.3$ and 0.65 dex, respectively.
Col.  (5) IR luminosity of the host galaxy from spectral energy distribution decomposition by \cite{Shangguan2018ApJ}.
Col.  (6) Axial ratio, derived from {\tt GALFIT} modeling of the host galaxy.
Col.  (7) References for the axial ratio.
Col.  (8) The inclination angle of the host galaxy.
Col.  (9) CO(1--0) line luminosity.  We convert the ALMA sample from $L^\prime_{\tiny \mbox{CO(2--1)}}$ to $L^\prime_{\tiny \mbox{CO(1--0)}}$ with a ratio of 0.62.
Col. (10) Molecular gas mass derived from CO line luminosity, assuming $\alpha_\mathrm{CO}=3.1\,M_\odot\,\mathrm{(K\,km\,s^{-1}\,pc^{2})^{-1}}$.
Col. (11) The width of the CO integrated profile at 50 percent of its maximum.
Col. (12) The morphology of the host galaxy: ``D'' = disk, ``E'' = elliptical, ``U'' = uncertain, ``M'' = merger, ``t'' = tidal disturbance feature, and ``c''  = companion.
Col. (13) References for the morphology.
\\
References: 
(1) \cite{Kim2017ApJS}; 
(2) Y. Zhao et al. (2020, in preparation); 
(3) \cite{Bentz2018ApJ};
(4) \cite{Zhang2016ApJ}; 
(5) \cite{Crenshaw2003AJ};
(6) \cite{Surace1998ApJ}.
}
\end{deluxetable*}

%% file: tab2.tex

\begin{deluxetable*}{c c c c r c c c}
\tablecaption{Outflow Properties of PG Quasars\label{tab:of}}
\tabletypesize{\scriptsize}
\tablehead{
\colhead{Object} &
\colhead{$\log\,L^\prime_\mathrm{CO(2\mbox{--}1), out}$} &
\colhead{$\log\,M_\mathrm{H_2, out}$} &
\colhead{$R_{\tiny \coline}$} &
\colhead{$\dot{M}_\mathrm{H_2, out}$} &
\colhead{$a_\mathrm{G}$} &
\colhead{$r_\mathrm{G}$} &
\colhead{$\chi^2_\mathrm{r}$}\\
\colhead{} &
\colhead{($\mathrm{K\,km\,s^{-1}\,pc^2}$)} &
\colhead{($M_\odot$)} &
\colhead{(kpc)} &
\colhead{($M_\odot\,\mathrm{yr^{-1}}$)} &
\colhead{($\arcsec$)} &
\colhead{} &
\colhead{}
}
\colnumbers
\startdata
 PG 0003$+$199 & $<6.68$ & $<6.79$ &                 1.05 &            $<57.23$ &  4.06$\pm$0.62 &              1 &    1.24 \\
 PG 0007$+$106 & $<7.72$ & $<7.83$ &                 3.25 &           $<203.04$ &  3.79$\pm$0.79 &              1 &    1.22 \\
 PG 0049$+$171 & \nodata & \nodata &              \nodata & \mcl{1}{c}{\nodata} &        \nodata &        \nodata & \nodata \\
 PG 0050$+$124 & $<7.29$ & $<7.40$ &                 2.23 &           $<109.70$ &  3.67$\pm$0.05 &  0.73$\pm$0.01 &    4.41 \\
 PG 0923$+$129 & $<6.71$ & $<6.82$ &                 1.54 &            $<41.37$ &  5.13$\pm$0.08 &  0.57$\pm$0.03 &    3.08 \\
 PG 0934$+$013 & $<7.33$ & $<7.44$ &                 2.00 &           $<133.36$ &  3.96$\pm$0.17 &              1 &    1.42 \\
 PG 1011$-$040 & $<7.54$ & $<7.65$ &                 1.84 &           $<237.18$ &  3.17$\pm$0.08 &              1 &    2.13 \\
 PG 1119$+$120 & $<7.26$ & $<7.37$ &                 1.25 &           $<184.41$ &  2.52$\pm$0.22 &              1 &    1.19 \\
 PG 1126$-$041 & $<7.30$ & $<7.41$ &                 4.12 &            $<60.32$ &  6.89$\pm$0.16 &  0.31$\pm$0.06 &    1.67 \\
 PG 1211$+$143 & $<7.73$ & $<7.84$ &                 2.60 &           $<257.20$ &  3.16$\pm$1.00 &              1 &    1.40 \\
 PG 1229$+$204 & $<7.44$ & $<7.55$ &                 6.54 &            $<53.15$ & 10.29$\pm$0.56 &  0.59$\pm$0.05 &    1.35 \\
 PG 1244$+$026 & $<7.29$ & $<7.40$ &                 0.78 &           $<314.63$ &  1.61$\pm$0.26 &              1 &    1.33 \\
 PG 1310$-$108 & $<6.92$ & $<7.03$ &                 1.84 &            $<55.94$ &  5.12$\pm$0.26 &              1 &    1.24 \\
 PG 1341$+$258 & \nodata & \nodata & \hspace{-1em}$<4.19$ & \mcl{1}{c}{\nodata} &        $<4.98$ &              1 &    1.15 \\
 PG 1351$+$236 & $<7.39$ & $<7.50$ &                 2.29 &           $<134.51$ &  4.14$\pm$0.13 &  0.75$\pm$0.05 &    1.59 \\
 PG 1404$+$226 & $<8.02$ & $<8.13$ &                 5.38 &           $<242.64$ &  5.75$\pm$0.78 &  0.43$\pm$0.14 &    1.20 \\
 PG 1426$+$015 & $<7.69$ & $<7.80$ &                 3.00 &           $<205.98$ &  3.61$\pm$0.21 &  0.67$\pm$0.15 &    1.22 \\
 PG 1448$+$273 & $<7.14$ & $<7.25$ &                 2.66 &            $<64.91$ &  4.14$\pm$0.22 &              1 &    1.20 \\
 PG 1501$+$106 & $<7.04$ & $<7.16$ &                 1.16 &           $<118.99$ &  3.15$\pm$0.85 &              1 &    1.14 \\
 PG 2130$+$099 & $<7.56$ & $<7.68$ &                 2.20 &           $<208.64$ &  3.62$\pm$0.24 &  0.72$\pm$0.06 &    1.30 \\
 PG 2209$+$184 & $<7.44$ & $<7.55$ &                 4.10 &            $<83.98$ &  5.95$\pm$0.21 &              1 &    1.27 \\
 PG 2214$+$139 & $<7.37$ & $<7.48$ &                 5.61 &            $<52.71$ &  8.46$\pm$1.05 &              1 &    1.06 \\
 PG 2304$+$042 & \nodata & \nodata &              \nodata & \mcl{1}{c}{\nodata} &        \nodata &        \nodata & \nodata \\
\enddata
\tablecomments{
Col. (1) Source name.
Col. (2) Upper limit of the CO(2--1) luminosity of the outflow.
Col. (3) Upper limit of the molecular gas mass of the outflow.  We adopt $R_{21}=0.62$ \citep{Shangguan2019ApJ} and $\alpha_\mathrm{CO}=0.8$ (e.g., \citealt{Cicone2014AA,Fiore2017AA}).  
Col. (4) The physical radius of the CO(2--1) line emission of the quasar host galaxy.  We adopt it as the upper limit of the outflow radius.
Col. (5) The mass outflow rate.
Col. (6) The major axis FWHM of the CO(2--1) line emission derived by fitting the uv data with the CASA task \texttt{uvmodelfit}.  We adopt a 3 $\sigma$ upper limit for PG~1341+258, whose line is too weak to be reliably detected.
Col. (7) The axis ratio of the elliptical Gaussian model of \texttt{uvmodelfit}.  If the data are not good enough to 
constrain the axis ratio, we adopt a circular Gaussian model (axis ratio fixed to 1).  When the elliptical Gaussian 
model is applicable, its best-fit major axis is not significantly different from that of the circular 
Gaussian model.
Col. (8) The reduced $\chi^2$ reported by \texttt{uvmodelfit}.
}
\end{deluxetable*}